\magnification1200

\rightline{KCL-MTH-05-09} \rightline{hep-th/0507152}

\vskip 1.5cm
\centerline
{\bf  Arbitrary Four String Scattering at High Energy and Fixed Angle}
\vskip .8cm
\centerline{Nicolas Moeller and Peter West}
\vskip .3cm
\centerline{King's College London}
\centerline{Department of Mathematics}
\centerline{London WC2R 2LS, UK}

\vskip 1.0cm
\leftline{\sl Abstract}
\vskip .2cm
\noindent
We calculate, using the group theoretic approach to string theory, the
tree and one loop scattering of four open and closed arbitrary bosonic
string states. In the limit of high energy, but fixed angle, the
multi-string vertex at tree and one loop levels that we find takes a
very simple form. We propose, and present arguments for, a form for
the high energy multi-string vertex at all loops; in particular we
give a path integral derivation of this vertex. Our results agree with
those of Gross and Mende for tachyon scattering amplitudes, but those
for any other string scattering are substantially different from that
discussed in reference [5]. We also develop some of the technology
used in the group theoretic method to compute loop corrections.

\vskip .5cm

\vfill
\eject
\medskip
{\bf {1. Introduction }}
\medskip

Long ago in the early days after the discovery of string theory,
high energy fixed angle scattering of the open string four tachyon
scattering was considered [1]. Extremising the known one loop
amplitude it was found that the non-planar process possessed a
saddle point and the amplitude was evaluated at that point. This
study was taken up in the late eighties  [2-5]. These authors
considered the closed string in the same limit and  found that the
amplitude for four tachyon scattering possesses a saddle point at
all genus  at which the positions of the Koba-Nielsen points were
fixed and whose cross ratio was related to the angle of
scattering.  Furthermore, a concise formula for four tachyon
scattering for any genus was derived   in the high energy, fixed
angle limit. An important role in this analysis was played by the
representation of the string world surface by the Riemann sphere
with two branch cuts whose degree was the genus. In reference [5]
it was proposed that there existed relations between any string
scattering amplitude and that for the same number of tachyon
scattering that was independent of the genus. The study of the
open string was also extended in reference [4].
\par
Some interesting papers have also considered high energy scattering stressing
the
role of the decoupling of zero norm physical states [6].
\par
In a theory that possesses a symmetry that is spontaneously
broken, the symmetry, although present at energy scales below  the
breaking scale, is obscured; however, above this scale the
amplitudes display the symmetry in a more transparent way.  As
such, it has been hoped that the study of string theory at high
energy will shed some light on the underlying symmetries. In more
recent times it has been proposed that M theory  should possess a
very large Kac-Moody algebra called $E_{11}$ [7]. This was
proposed by studying the maximal supergravity theories  that
encode all perturbative and non-perturbative effects of string
theory at low energy. The motivation for this paper  was to  study
the
  results of references [1-5] to see if these  can make contact with
this proposed large symmetry.
\par
We will study the high energy, fixed angle,  scattering of
arbitrary physical bosonic string states at tree and one loop
level. The most efficient method to  study arbitrary string
scattering is to use multi-string scattering vertices. These
emerged initially from
  factorizing tachyon  string scattering amplitudes, and
was carried out in the earliest days of string theory. By sewing such
vertices the pioneers of string theory were able to calculate
multi-string amplitudes, so lying the foundations for string theory. In
particular, they computed the one loop open string scattering amplitudes
[26-32] and developed a projector which was used to  eliminate the ghosts
for this amplitude [33]. For
two  reviews which set out some of this remarkable progress, see
reference [8]. These results were found before our present understanding
of the systematic BRST techniques to  add ghosts and so the old results
often suffered from the presence of spurious ghosts. Their other problem
was that the multi-string vertices they found emerged from factorising
tachyon amplitudes and their properties were not well understood.
\par
In the late eighties the study of multi-string vertices was taken up
again in the so-called group theoretic approach [9-22]. The essence of
this approach was that the multi-string vertices were the simplest
objects in string theory.  It was found that multi-string vertices
satisfied some very simple properties called overlap relations, and
these were then formulated in such a way that they were used to define
what a multi-string vertex was. This allowed for a systematic
construction, and elaboration of the properties, of multi-string
vertices.  For example, it allowed the incorporation of ghost
oscillators, first given in an earlier paper [21], and a systematic
method to change the dependence of the vertex on the Koba-Nielsen
points and moduli. The amplitude was found by saturating the
multi-string vertex with the desired physical states which were to be
scattered and integrating, with an appropriate measure, over the
Koba-Nielsen points and moduli of the Riemann surface swept out by the
string. As we have mentioned, the vertex is determined by certain very
simple overlap relations while the measure is determined by demanding
that zero norm physical states decouple.  Using this technique, the
scattering amplitudes for the open bosonic string [10,12-14],
including ghosts [15,16,21], and superstring [10,19,20] were
calculated. The tree level scattering for the closed string was found,
but this string was not as systematically investigated as the open
string. We rectify this omission in this paper. The group theoretic
method was developed in the rather long series of papers [9-22].
Although the early papers contained the essential steps, they often
lacked some of the insights that were gained later on, and so they are
not always so easy to read. A review was given in reference [18], but
we give an account of the group theoretic method in section two of
this paper in such a way that the rest of the paper can be understood
in a self-contained manner. Some aspects of the group theoretic
approach, such as the integrated overlap identities, were considered
in the approach of reference [23] which origined [25] in the
properties of the sum over world-sheets approach to string theory.
\par
In section three we use the group theoretic method to calculate the high
energy, fixed angle, behaviour of open and closed bosonic strings
scattering  at tree level, while in section four we calculate
four open and closed bosonic string  scattering at one loop.
The multi-string vertices we find in
the high energy limit do have an extremely simple form, that for one
loop closed string scattering is given by
$$
V_c=\prod_{k=1}^N {}_{(k)}\langle
0|\exp\left(-\sum_{i,j}\sum_{n=1}^{\infty}\left( \alpha_{2n}^{ \mu
(i)}{1\over 2n(\varsigma_j-\varsigma_i)^n}\alpha_{0 \mu}^{(j)}
+\bar \alpha_{2n}^{ \mu (i)}{1\over
2n(\bar\varsigma_j-\bar\varsigma_i)^n}\alpha_{0 \mu}^{(j)}\right)
\right.
$$
$$ \left. +{}\sum_{i,j}
\alpha_0^{ \mu (i)}\ln \sqrt{|(
\varsigma_j-\varsigma_i)|}\alpha_{0 \mu}^{(j)}\right) \eqno(1.1)$$
where $\varsigma_j$ are the positions of the Koba-Nielsen points,
which are fixed to be at the beginning and end of the two branch
cuts on the Riemann sphere  that represents the string
world-sheet. We conjecture, with supporting evidence, the results
for arbitrary genus in section seven.
\par
In sections five and six we further develop some of the aspects of the
group theoretic method associated with the calculation of loop string
scattering.
\par
Our results for four tachyon scattering in the high energy,
fixed angle,  limit agree with those of references [1-4] and so we agree
with the majority of the work in these papers. However, we find that our
results for physical states other than tachyons differ substantially from
those of reference [5].  Indeed,  we
find  that some of the amplitudes which are non-zero  in
reference [5]  vanish in the extreme high energy limit. As a result, we do
not find the all genus symmetry relations found in reference [5]. At
the end of section four we pinpoint where we believe the problems arise
in reference [5].

\bigskip
{\bf {2. Review of the group theoretic approach}}
\medskip
Here we  review the group theoretic approach to string theory,
which was developed in references  [9-20]. The following results
will be valid for both open and closed bosonic strings, and with
suitable extensions the superstrings.  As $N$ strings scatter they
sweep out a world-sheet which is a two dimensional surface of
genus $g$ with some marked points, or punctures, where the
outgoing strings leave the surface. We refer to these points as
Koba-Nielsen points. The string theory possesses certain conformal
operators, denoted generically by the symbol $R(z)$, which depend
on the world-surface, and act on the string Hilbert space  ${\cal
H}$. By definition a conformal operator of weight $d$ behaves
under a conformal transformation $\xi\rightarrow \xi^\prime(\xi)$,
where $\xi$ is a local coordinate on the world-surface, as
$$
R(\xi) \rightarrow R^\prime(\xi^\prime)=
U R(\xi)   U^{-1}= R(\xi)\left({d
\xi \over d \xi'}
\right)^d .
\eqno(2.1)$$
Here $U$ is the operator on the string Hilbert space that corresponds to
the conformal transformation $z\rightarrow z^\prime(z)$. The operator
$U$ can be written locally in the form
$$
U=\exp\left(\sum_{n=-\infty}^{\infty} a_n L_n\right)
\eqno(2.2)$$
For the open string we only
consider operators of this type, but for closed strings conformal
operators have left and right weights
$(d,\bar d)$,  but we will mainly only require operators with a weight
which is either
$(d,0)$ as above or $(0,\bar d)$ in which case we will have the conformal
transformation on the coordinate $\bar z$ be implemented by $\bar
L_n$.
For example, for the open bosonic string, the conformal operator is given by
$\partial Q^\mu$ which has conformal weight one. Equation
(2.1) also holds for the embedding field itself $Q^\mu$  with weight zero.
 The energy momentum tensor $T(z)$ transforms with weight two, but
also has an anomalous term which involves the Schwarzian derivative
of  the conformal transformation. In particular, it obeys
$T^\prime(z^\prime)=T(z)\left({d z
\over d z'}\right)^2+{c\over 12} (Sz)(z')$
where $(Sf)(z)={f^{\prime\prime\prime}(z)\over f^{\prime}(z)}-{3\over
2}\left({f^{\prime\prime}(z)\over f^{\prime}(z)}\right)^2$ and $c$ is the
central charge associated with $T$.     We also have the ghosts
$b$ and
$c$ of conformal weights two and $-1$ respectively.
\par
In
the group theoretic approach, the scattering amplitude, at genus
$g$, of $N$ arbitrary on-shell physical states $|\chi_i\rangle$ is given
in terms of a vertex $V$, by the expression
$$
A(|\chi_1\rangle, |\chi_2\rangle, \ldots, |\chi_N\rangle) = \int_{{\cal
M}_{g,z_k}} \Pi_r dv_r \Pi^\prime_i dz_i
{f_m}(z_i, v_r) V |\chi_1\rangle_{(1)} |\chi_2\rangle_{(2)} \ldots
|\chi_N\rangle_{(N)}.
\eqno(2.3)$$
Here ${\cal M}_{g,z_k}$ is the moduli space of Riemann surfaces of
genus
$g$ with  $N$ marked   points $z_i$, i.e the
Koba-Nielsen points.  The integration runs over both the $z_i$  and over
the  moduli $v_r$. The ways of representing the two dimensional surface
will be discussed later. The  function
${f_m}(z_i, v_r)$ is the measure, and
 $V$ is what we call the
{\it vertex}. It is a multi-linear map from $N$ copies of the Hilbert
space of the string
into the complex numbers:
$$
V \ : \ {\cal H}_1 \otimes \ldots \otimes {\cal H}_N
\rightarrow {\bf C}.
\eqno(2.4)$$
It depends on the Koba-Nielsen variables and on the moduli. The same
vertex provides the scattering for any physical states once it is
saturated with the corresponding physical states as in equation (2.3).
Thus the vertex   does not depend on what physical states are scattering,
and therefore once we know the vertex and the measure, all scattering
amplitudes at genus $g$ are treated on the same footing. We will now
define what is meant by the vertex and show how it and the measure are
determined.
\par
The string vertex $V$ is defined by demanding that it satisfy the overlap
equations
$$
V R^{(i)}(\xi_i) = V R^{(j)}(\xi_j) \left({d \xi_j \over d \xi_i}
\right)^d. \eqno(2.5)$$ Here $\xi_i$ and $\xi_j$ denote coordinate
patches on the string world-surface that are valid in the
neighbourhood of the Koba-Nielsen points $z_i$ and $z_j$
respectively. The coordinate $\xi_i$ associated with the
Koba-Nielsen point $z_i$  is required to vanish at this  point,
i.e. $\xi_i(z_i)=0$, and similarly $\xi_j(z_j)=0$. Furthermore,
the two coordinates are related in their overlap region by an
analytic transformation which we denote  by
$\xi_j=\tau_{ji}(\xi_i)$. In equation (2.5) it is important to
realise that the conformal operator $R^{(j)}$ of weight $d$ acts
on the $j$th leg, or string Hilbert space of the vertex. This
equation states that when the action of two conformal operators on
the vertex are defined in an overlap region then each operator
gives the same result when referred to the same coordinate system.
Equation (2.5) holds for the conformal operators used to construct
the string under consideration. For the closed bosonic string
these include $x^\mu$ and $\partial x^\mu$ and the energy momentum
tensor $T$ provided one includes its anomalous term. Of course,
some overlap identities
 can be derived as a consequence of some others. The
obvious example being that $\partial x^\mu$ follows from the identity for
$x^\mu$ by differentiation of the overlap equation.
At first sight it may appear that equation (2.5) just relates the
conformal properties of an operator $R$ of weight $d$, however, the
operators
$R$ on the left and right hand side of the equation act on different
one-string Hilbert spaces in a highly non-trivial manner.
\par
One is free to
choose whatever coordinate system one likes provided it obeys the above
criterion, and one finds correspondingly different string vertices.
In fact the vertex only depends on the transition functions $\tau_{ji}$
between the coordinate patches rather than the local coordinates
themselves as only the former quantities actually enter into the overlap
equations. To see this we  write $\xi_j$ in equation (2.5) as
$\tau_{ji}(\xi_i)$ and we note  that $\xi_i$ plays the role of a dummy
variable.  It is often useful to refer the
coordinates
$\xi_j$ to some underlying coordinate $z$, one choice of which  we
will discuss shortly, and then we may write
$$
\xi_i(z)= \sum_{n=1}^\infty{a_n^i (z-z_i)^n} \equiv
(V_i)^{-1}(z)\quad ,
\qquad a_1^i \neq 0.
\eqno(2.6)$$
In terms of the  coordinate $z$, the Koba-Nielsen points are  denoted
by  $z=z_i$, $i=1,\ldots, N$ and it has the above form as
$\xi_i(z_i)=0$. We note that the transition functions are given by
$\tau_{ji}= (V_j)^{-1} V_i$. The choice of any  local coordinates
$\xi_i$, subject to the above conditions,  corresponds to taking any
values of the constants
$a_n$.  However, as we will see  different choices of local
coordinates  lead to the same physical scattering amplitudes.
\par
The expression $V R^{(i)}(\xi_i)$ can be expressed as a Laurent series in
$\xi_i$ around $\xi_i=0$, but when $\xi_i$ approaches another
Koba-Nielsen point, say $z_j$, this series diverges in general. To obtain
a convergent expression as one approaches $z_j$ one uses the overlap
identity of equation (2.5) in the overlap region where both  $V
R^{(i)}(\xi_i)$ and  $V R^{(j)}(\xi_j)$ are convergent, and then
approaches the point $z_j$ using the latter expression. As such,  the
domain of validity of the coordinate $\xi_i$ is defined in this sense of
convergent series rather than in the sense of coordinate patches as used
in  the theory of differentiable manifolds.
\par
For scattering at genus zero the world-sheet is just the Riemann
sphere with the marked Koba-Nielsen points and  we can introduce a
coordinate $z$ which covers the whole sphere with the exception of
the point at infinity where we use the coordinate $\zeta={1\over
z}$. If none of the Koba-Nielsen points are at infinity, one can
use the local coordinates $\xi_i=z-z_i$, but there are of course
an infinite number of other choices. Should it happen that one has
a Koba-Nielsen point at infinity then one can use the local
coordinate $\zeta$ for that string.
\par
For surfaces of genus greater than zero the string world-surface
is a Riemann surface with marked points and handles. We may use
either the Schottky or Fuchsian representation of the surface. In
the Fuchsian representation of the Riemann surface, every Riemann
surface of genus $g \geq 2$ is conformally equivalent to the
quotient of the upper-half plane by a fixed point free Fuchsian
group (a discontinuous subgroup of ${\rm PSL}(2, R)$). While a
genus one surface can be mapped to a parallelogram on the complex
plane whose opposing sides are identified. Hence, for all genus
surfaces we need only one   coordinate $z$, with certain
identifications
$$
z \sim P_n z,\ n=1,\ldots ,2g \eqno(2.7)$$ where $P_n$ are
generators of the Fuchsian group if $g \geq 2$, or in the case of
genus one the two translations that identify the opposing sides of
the parallelogram. When considering the open string  we will
double the open string world-surface to work with a more
conventional Riemann surface.
\par
We now
demand that the vertex, by definition,  satisfy the  further overlap
conditions, namely
$$
V R^{(i)}(\xi_i) = V R^{(i)}(P_n^i \xi_i) \left({d (P_n^i \xi_i)
\over d
\xi_i} \right)^d,\ \ n=1,\ldots ,2g
\eqno(2.8)$$
Essentially this equation just states that $V R$ is defined on the actual
surface.  Using equation (2.5) to change equation (2.8) from leg $i$ to leg $j$
we
find that
$$
P_n^j=\tau_{ji} P_n^i(\tau_{ji})^{-1}\ \ \ \ {\rm or}\ \  \ \ V_i
P_n^i (V_i)^{-1} = V_j P_n^j (V_j)^{-1} \equiv P_n.
\eqno(2.9)$$
In carrying out this
step we have expressed $\xi_i$ as $\xi_i=\tau_{ij}\xi_j$,  the
fact that  the argument $P_n^i\xi_i$ of
$R^{(i)}$ changes using the overlap to  $R^{(j)}$ with argument
$\tau_{ji}(P_n^i\xi_i)$ and that $(\tau_{ji})^{-1}=\tau_{ij}$.
In the last part of equation (2.9) we have used the expression for the
transition functions in terms of the $V_i$ and as a result
$$P_n^i=  (V_i)^{-1} P_n V_i.
\eqno(2.10)$$
We recognise $P_n$ as the identification expressed in terms of the $z$
coordinate system as in equation (2.7). The story for the Schottky
representation of the surface is similar, but one has fewer overlap
conditions as one has fewer identifications. The reader is referred to
reference [13] for an account.
\par
Hence, the vertex $V$ is just defined to satisfy equation (2.5)
and, in the case of loop corrections,  equation (2.8). What is far
from obvious, but  true, is that these two equations determine the
string vertex $V$ up to a  function of the Koba-Nielsen
coordinates $z_i$ and moduli $v_r$. One may choose this function
by taking $V\prod _i |0\rangle_{(i)}=1$, but any other choice of
the function is acceptable and will be taken into account when one
computes the measure.
\par
The amplitude of equation (2.3) also involves the measure $f$ which
is determined from the requirement that
the amplitude should vanish if any of the
$|\chi_i\rangle$ is a zero norm physical state. This means that
$$
A(|\chi_1\rangle, |\chi_2\rangle, \ldots, |\chi_N\rangle) = 0
\eqno(2.11)$$
if any of the $|\chi_i\rangle$ are physical zero norm states. This
requirement is similar to the decoupling of such states
in quantum electrodynamics and so we can think of equation (2.11)
 as a kind of Ward identity. In fact, this fixes the measure $f_m$ up to a
constant. The constant is determined as in the old days of sewing,
or the sum over world-surfaces approach, by requiring that the
scattering amplitude is unitary.
\par
{\bf To summarize the group theoretic approach, the string vertex $V$ is
by definition an object which obeys the overlap equations (2.5) and (2.8)
corresponding to the conformal operators used to define the string
theory, the string scattering amplitude is given by equation (2.3) and the
measure is determined by demanding that zero norm physical states
decouple. This is the complete system required to find the scattering of
arbitrary string states}.
\par
Most of the vertex can be found more easily from the so called integrated
overlap identities which we now  derive from the unintegrated overlap
identities above. Let us consider the expression
$$
V \left(  {\oint_{0}{d\xi_i \varphi R^{(i)}(\xi_i)}} \right)
\eqno(2.12)$$
where $\varphi$ is a meromorphic tensor of weight $1-d$,
that is under a conformal transformation it transforms as
$$
\varphi(\xi) \to \varphi^{\prime}(\xi^\prime) =\varphi(\xi)
\left({d \xi \over d \xi'} \right)^{1-d} . \eqno(2.13)$$ As a
result, $V\varphi R^{(i)}(\xi_i)$ is a tensor of weight one, or
simply, a one form. Let us now consider deforming the contour
around $z_i$ such that it encloses any other non-analytic
behaviour of $V \varphi R^{(i)}(\xi_i)$.  We take an arbitrary
$\varphi $ of weight $1-d$ provided it is analytic except for
poles around the other Koba-Nielsen points. Thus we find a set of
contour integrals around the Koba-Nielsen points. In carrying out
this contour deformation one
 approaches the neighbourhood of other  Koba-Nielsen points and, as
explained above,  one must use the overlap equation (2.5) to maintain
convergence. Thus the overlap relations are the essential ingredient in
the derivation.  For the case of genus zero the contour is pulled off to
infinity where it vanishes, provided this is not a Koba-Nielsen
point, while at higher genus we find a contour around the  boundary of the
region which represents the Riemann surface. Since
$\varphi$ is defined on the surface, that is  it is suitably periodic
meaning
$$
\varphi(\xi_i) = \varphi(P_n^i\xi_i) \left({d (P_n^i\xi_i) \over d
\xi_i} \right)^{1-d},
\eqno(2.14)
$$
this contour vanishes using
equation (2.8).  Thus we find the integrated overlap equation
$$
\sum _jV \left( {\oint_{0}{d\xi_j R^{(j)}(\xi_j)} \varphi
\left({d\xi_j\over d\xi_i}\right)^{d-1}}\right)=0
\eqno(2.15)
$$
where we have kept explicit the transformation of $\varphi$ under a 
change of coordinates by writing the derivative term in front of $\varphi$;
this is consistent with the previous literature on the group theoretic approach.
We could also have written 
$\sum _jV \left( {\oint_{0}{d\xi_j R^{(j)}(\xi_j)} \varphi(\xi_j)}\right)=0$, 
but then we have to remember that $\varphi$ is a tensor that transforms as 
in (2.13).

\par
The vertex by  definition satisfies the overlap equations (2.5) and (2.8),
and  to discover its properties it is often simpler to find how these
equations behave under the desired transformation than to look into the
details of the vertex itself. In particular we now consider the effect
of   conformal transformations on the vertex. Let us consider a conformal
transformation
$\xi\to {\cal N}(\xi)$ and  denote by the same symbol
$ {\cal N}$ the operator of the realisation of the same conformal
transformation on the string Fock space. The reader can distinguish from
the context which $ {\cal N}$  is required. We can carry out different
conformal transformations on the different string Hilbert spaces, and as
before the superscript indicates which Hilbert space the operator is
acting on. For instance, $ {\cal N}^{(i)}$ acts on the $i$th Hilbert
space. We can define a new vertex by
$$  V(\hat z_i, \hat v_r, \hat \xi_i)=V( z_i, v_r,\xi_i) \Pi_k
{\cal N}^{(k)}_k
\eqno(2.16)$$
The action of the conformal transformation may change the
Koba-Nielsen points $z_i$,  the moduli $v_r$ and the local coordinates
$\xi_i$ used to define the vertex. We denote the new quantities with hats
and show the explicit dependence of the vertex on them.  The
subscript on
$ {\cal N}^{(k)}_k$ indicates the possibility to carry out
different conformal transformations on different string Hilbert spaces or
string legs. In general the operator ${\cal N}$ has the form
$$
 {\cal N}^{(k)}_k=\exp\left(\sum_{n=-\infty}^{\infty} c_n^k L_n^{(k)}\right)
\eqno(2.17)$$
where $c_n^k$ are arbitrary constants.
Under the above conformal transformation the vertex may also scale
  by a function of $v_r$ and $z_j$ which we have not explicitly
indicated. As mentioned above, the behaviour of
the vertex under a conformal transformation can be understood by studying
its effect on the overlap equations. Using the properties of conformal
operators of  equation (2.1)  we find that the overlap of equation (2.5)
becomes
$$
V(\hat z_i, \hat v_r, \hat \xi_i) R^{(i)}({\cal N}_i^{-1}(\xi_i)) =
V(\hat z_i, \hat v_r, \hat \xi_i)
R^{(j)}({\cal N}_j^{-1}(\xi_j)) \left({d ({\cal
N}_j^{-1}(\xi_j))
\over d ({\cal N}_i^{-1}(\xi_i))}
\right)^d.
\eqno(2.18)$$
Writing $\xi_j=\tau_{ji}(\xi_i)$ and making the change
in the dummy variable $({\cal N}^{(i)}_i)^{-1}\xi_i$ to $\hat \xi_i$
we find that the overlap relation becomes
$$
V(\hat z_i, \hat v_r, \hat \xi_i) R^{(i)}(\hat \xi_i) = V(\hat z_i, \hat
v_r, \hat \xi_i) R^{(j)}({\cal N}_j^{-1}\tau_{ji} {\cal
N}_i(\hat \xi_i))
\left({d ({\cal N}_j^{-1}\tau_{ji} {\cal N}_i(\hat \xi_i))
\over d \hat \xi_i}
\right)^d.
\eqno(2.19)$$
Interpreting the argument of the operator $R$ on the right hand side of
this equation as the new local coordinate, $\hat \xi_j=\hat \tau_{ji}(\hat
\xi_i)$, we find that the new local coordinates  and Koba-Nielsen points
of the vertex after the conformal transformation are given by the
equation
$$\hat \tau_{ji}=({\cal N}_j)^{-1}\tau_{ji}{\cal N}_i
 \eqno(2.20)$$
for all $i,j=1,\ldots , N$. The Koba-Nielsen points are encoded as they
are the points where $\hat \xi_j=0$ for each $j$.  Carrying out a similar
computation on the overlap equation (2.8), we conclude that the moduli of
the vertex, which are encoded in the identifications $P_n^i$, after the
conformal transformation are given by the equation
$$
\hat P_n^i= ({\cal N}_i)^{-1} P_n^i {\cal N}_i, \ i=1,\ldots ,N
 \eqno(2.21)$$
\par
Although one can work entirely in terms of the overlap functions,
it is advantageous to work in terms of the coordinate $z$ discussed
above. To this end we use
$\tau_{ji}=(V_j)^{-1}V_i$ and then equation (2.20) becomes
$$
V_j {\cal N}_j(\hat V_j)^{-1} =V_i{\cal N}_i(\hat V_i)^{-1}\equiv {\cal N}
\ \ {\rm or }\ \ {\cal N}_i=( V_i)^{-1}{\cal N}\hat V_i
\eqno(2.22)$$
We recognise ${\cal N}$ as a  conformal transformation  acting on the
$z$ coordinate system.
Using equations (2.9) and (2.22), we find that equation (2.21) can be
written as
$$
\hat P_n=  ({\cal N})^{-1} P_n  {\cal N}
\eqno(2.23)$$
\par
Let us  consider  a conformal transformation which
 leaves the moduli $v_r$ inert,  but  changes the positions of the
Koba-Nielsen points
 and the  local
coordinate systems $\xi_i\to \hat \xi_i=(\hat V_i)^{-1}z$ around
each Koba-Nielsen point. From equation (2.23) such a
transformation will have $P_n=\hat P_n$  and so we may take ${\cal
N}=I$ and, from equation (2.22), have ${\cal N}_i=( V_i)^{-1}\hat
V_i$. Acting on the vertex we have
$$
V(v_r, \hat{z}_j,\hat \xi_j) = V(v_r, z_j,\xi_j) \prod_{i=1}^N(V_i)^{-1}
\hat{V}_i.
\eqno(2.24)
$$
The new Koba-Nielsen points are given by the points where $\hat \xi_j$
vanish for each $j$. This is $\hat \xi_j(\hat z_j)=0$,  or equivalently
$(\hat V_j)^{-1}(\hat z_j)=0$.
Transformations ${\cal N}_i$ of this kind are all analytic
transformations.  Those for which ${\cal N}_i(0)=0$ do not change the
positions of the Koba-Nielsen points and just define new local
coordinates around each  Koba-Nielsen point. These are of the form of
 (2.17) with the sum  for $n\ge 0$. Those that also shift the
Koba-Nielsen points also include $L_{-1}$ in the sum.
 Since physical states are annihilated by
$L_n-\delta_{n,0}, n\ge0$ the former transformations   do not affect
physical amplitudes except for a scaling due to the $L_0$ term. However,
the   $L_0$ term may scale the vertex by a factor,  but this is
compensated for  when the measure is calculated.  The same
conclusion applies to the transformations that shift the Koba-Nielsen
points.
\par
Now we consider  an infinitesimal conformal transformation which
 does not change $z_j$, or the local coordinates, but  changes the moduli
$v_r \rightarrow \hat{v}_r = v_r + \epsilon h_r(v)$. In this case,
$V_j =\hat{V}_j$,  and we are left with ${\cal N}$. Then
equation  (2.23) applied to the coordinate $z$ becomes
$$
P_n(z) + \epsilon f(z) {d P_n(z) \over dz} = \hat{P}_n(z) + \epsilon f(P_n
z)
\eqno(2.25)
$$
where
$$
{\cal N}(z) = z + \epsilon f(z).
\eqno(2.26) $$
An infinitesimal  conformal transformation $z\to z + \epsilon f(z)$ is
realised on the string Hilbert space by the operator
$I+ \epsilon \oint_0  T(z) f(z)$. Consequently, implementing the conformal
transformation of equation (2.26) on the vertex we find that equation
(2.16) is given by
$$
\sum_{r=1}^g h_r(v) {\partial V \over \partial v_r} = V \left(
\sum_{j=1}^N
\oint_{\xi_j = 0} d \xi_j {T^{(j)}(\xi_j) } f^j(\xi_j) + c_f
\right)
\eqno(2.27)
$$
where $c_f$ is a function of the moduli and Koba-Nielsen points and
$\xi_i+ \epsilon f^i(\xi_i)= {\cal N}_i (\xi_i)= ( V_i^{-1}{\cal N}
V_i)(\xi_i)$.
The constant $c_f$ is required to take account of the
presence of a multiplier which has
not been explicitly written in equation (2.16) and the fact that
the vertex is only determined up to a function of the moduli and
Koba-Nielsen points by the overlap relations. The value of
$c_f$ is
easily determined
by taking a suitable state, such as the vacuum state,  on either side of
this equation. We recall,
however, that to find the actual scattering amplitude, we multiply the
vertex by the measure
$f_m$ and integrate over all the Koba-Nielsen variables and moduli.
The final result is independent of what choice of
multiplicative factor in front of  $V$ one takes.
\par
It is instructive to consider conformal transformations that do not
change the vertex, that is  they   change neither $v_r$,   $z_j$, nor the
local coordinate system. When we have loops this means that $P_n=\hat P_n$
and so we require
$$
P_n^{-1} {\cal N} P_n = {\cal N}.
\eqno(2.28)
$$
If there are no loops then there are no identifications $P_n$ and so
${\cal N}$ is unrestricted. The conformal transformations on each leg
are given by equation (2.22) with $V_i=\hat V_i$ and so
$$£{\cal N}_i= ( V_i)^{-1}{\cal N} V_i
\eqno(2.29)
$$
For infinitesimal transformations, we take equation (2.25), with $P_n =
\hat{P}_n$, which means that $f(z)$ transforms as a vector field
(i.e., a tensor of rank $-1$) and is defined on the Riemann surface
itself. Thus we find an equation of the form
$$
V \left\{ \sum_{j=1}^N \oint_{\xi_j=0} d\xi_j \varphi \left( T^j(\xi_j)
 {d\xi_j \over d\xi_i} +{ c\over 12}(S\xi_j)(\xi_i){d\xi_i \over d\xi_j}
\right)
\right\} = 0
\eqno(2.30)
$$
This is just an example of the overlap equation (2.5) found by taking
$R(z) = L(z)$ which has conformal weight 2 and we have recognised the
constant to be the Schwarzian derivative.  The set of such
transformations which preserve a   $g$ loop vertex with
$N$ external lines form a group which is discussed in reference [17].
\par
 Under the transformations
$$
 V_j\to S V_j, \ \ \ P_n\to S P_n S^{-1}
\eqno(2.31)$$
the vertex is inert as it only depends on $P_n^i$ of equation (2.10) and
the transition functions $\tau_{ji}= (V_j)^{-1}
(V_i)$ and  since these are inert under the above transformations the
result follows. Indeed, the transformation properties of the vertex can be
summarized by writing the dependence of the vertex $V$ on $v_r$ and
$z_j$, $\xi_j$ through $P_n$ and $V_j$, i.e.  $V(P_n, V_j)$ and the
${\rm SU}(1,1)$ invariance becomes $ V(S P_n S^{-1}, S V_j) = V(P_n, V_j)$.
This invariance is a feature of the redundancy of the representation in
terms of $P_n$ and $ V_j$ that we have used.
The conformal transformation property of the vertex embodied in
equations (2.22) and (2.23) may be  written as
$$
V(P_n, V_j) \prod_{i=1}^N{\cal N}^{(i)}_i = V({\cal N}^{-1} P_n
{\cal N}, \hat{V}_j) . \eqno(2.32)
$$
We have again not explicitly written  the possible scaling of $V$ by a
function of $v_r$ and $z_j$.
\par
As stated above, the measure is determined by ensuring that zero-norm
physical states decouple. For the open bosonic string
these states are of the form
$$
L_{-1}|\Omega\rangle \quad , \quad \left(L_{-2} + {3\over2} L_{-1}^2
\right)
|\Omega^\prime\rangle
\eqno(2.33)$$
where
$$
L_n|\Omega\rangle = L_n|\Omega'\rangle = 0 \quad n > 0,\ \ \ \ \ \
L_0|\Omega\rangle = (L_0+1)|\Omega'\rangle = 0 \eqno(2.34)$$
together with higher level zero norm states. The states for other
strings have a similar form; they consist of operators which
generate the symmetry algebra of the string in question acting on
a state which is highest weight with respect to the algebra.
Saturating the vertex with one of these states we may interpret it
as an infinitesimal conformal transformation  applied to the
vertex times $|\Omega'\rangle$ or $|\Omega\rangle$. Such
transformations will either preserve the vertex, or change its
Koba-Nielsen coordinates  or moduli. Changes of local coordinate
systems are irrelevant as the vertex is saturated with physical
states. Consequently, we find, generically, that
$$
\int {\prod_k}^\prime dz_k \prod_r dv_r {f_m}
\left(\sum_r h_r {\partial V \over \partial v_r} + \sum_j g_j
{\partial V \over \partial z_j} - c V \right)
|\chi_1\rangle_{(1)} \ldots |\Omega\rangle_{(i)} \ldots
|\chi_N\rangle_{(N)} = 0
\eqno(2.35)
$$
where all the states $|\chi\rangle_{(i)}$  are physical and so
obey $(L_n-\delta_{n,0})|\chi\rangle = 0$, $n \geq 0$ and
$$
v_r \rightarrow v_r + h_r(v_r, z_j),\ \ \ \ \
z_j \rightarrow z_j + g_j(z_j, v_r)
\eqno(2.36)$$
are the changes that the infinitesimal conformal transformation
induces on the vertex. We find, by integrating by parts, that
$$
0 = \sum_r{\partial \over \partial v_r} (h_r  {f_m}) +
\sum_i{\partial
\over \partial z_i} (g_i {f_m}) + cf_m
\eqno(2.37)
$$
in other words a first-order differential equation for the measure,
${f_m}$. In fact, one has as many differential equations as one has
Koba-Nielsen variables and moduli and so the measure is determined by
them.  In this calculation  we have neglected boundary terms which
in general are not zero and correspond to  a B.R.S.T anomaly. Indeed, all
open strings, except for the gauge group ${\rm SO}(2^{13})$, have such an
anomaly [12,25].
\par
The process of decoupling without ghosts becomes, at least at
first sight, more and more difficult to implement at higher loops due
to the complicated structure of the zero-norm physical states. This
difficulty is not present when one incorporates ghosts as these states
are of the form $Q|\Lambda\rangle$. However, the correct vertex to use
is no longer $V$, when ghost extended, as this is annihilated by $Q$
and so does not lead to any constraints on the measure. In fact, one
must multiply the vertex by a ghost oscillator prefactor. This is done
explicitly in references [15] and [16] and one finds that the measure
obeys the differential equations
$$
{\partial \ln {f_m} \over \partial z^j} = 0 = {\partial \ln {f_m}
\over \partial v_r} + c_r
\eqno(2.38)
$$
where $c_r$ are the constants in equation (2.27). One can, of course,
redefine one's vertex by multiplying it by ${f_m}$ and then $c_r$ is
zero  and the corresponding measure is one.
\par
This completes our review of the group  theoretic method to computing
string amplitudes. To illustrate how to apply the method, we compute
open bosonic string scattering at the tree level. We adopt a
particularly simple set of local coordinates
$$\xi_i(z)=(V_i)^{-1}(z)=z-z_i, \ \ {\rm or }\ \ \xi_j=\xi_i+z_i-z_j
\eqno(2.39)$$
We begin by deriving the
$N$ string vertex
$V$. The overlap equation (2.5) for the conformal operator
$$
Q^{\mu}(z) = -{\overleftarrow{\partial} \over \partial \alpha_0^\mu}
 - \alpha_0^\mu \ln z
+\sum_{{n=-\infty}\atop{n\neq 0}}^\infty {\alpha_n^\mu z^{-n}
\over n}
\eqno(2.40)
$$
of conformal weight zero is given by
$$
V  Q^{\mu (i)}(z-z_i) =V Q^{\mu (j)}(z-z_j) .
\eqno(2.41)
$$
For $R(z) = P^\mu(z)\equiv\partial
Q^\mu(z)$, the integrated
overlap equation (2.15) becomes
$$
V \left( \sum_{j=1}^N \oint {d \xi_j } \varphi_n
P^{\mu (j)}(\xi_j) \right) = 0.
\eqno(2.42)
$$
We can take $\varphi_n$ to be of the form
$$
\varphi_n = {1 \over (\xi_i)^n} = {1 \over (\xi_j + z_j - z_i)^n} =
\sum_{p=0}^\infty \left(\matrix{-n \cr p}\right) {(\xi_j)^p \over
(z_j-z_i)^{n+p}} \qquad \forall\ \  j \neq i.
\eqno(2.43)
$$
In fact, this set of $\varphi$'s form a complete set of all
allowed $\varphi$'s. Substituting the $\varphi$'s into equation (2.42)
and using the relation
$$
P^\mu(z) = \sum_{n=-\infty}^\infty \alpha_{-n}^\mu z^{n-1}
\eqno(2.44)$$
we find the identity
$$
0 = V \left( \alpha_{-n}^{\mu (i)} + \sum_{{j=1}\atop{j \neq i}}^N
\sum_{p=0}^\infty \left(\matrix{-n \cr p}\right) {\alpha_p^{\mu (j)} \over
(z_j-z_i)^{n+p}} \right).
\eqno(2.45)
$$
The solution to this equation gives us the vertex for the open string
$$
V = \prod_{k=1}^N {}_{(k)}\langle 0|
\exp\left(\sum_{{i,j}\atop{i\neq j}} \left\{ -{1\over2}
\sum_{n,m=1}^\infty \alpha_n^{\mu (i)} {(n+m-1)! (-1)^m \over m!
n! (z_j-z_i)^{n+m}} \alpha_{m\mu}^{(j)} \right. \right.
$$
$$
\left. \left. -\sum_{n=1}^\infty{\alpha_n^{\mu (i)}
\alpha_{0\mu}^{(j)} \over n (z_j-z_i)^n} - {1 \over 2}
\alpha_0^{\mu (i)} N^{ij} \alpha_{0\mu}^{(j)} \right\}
\matrix{{}\cr {} \cr {}} \right) \eqno(2.46)
$$
where $N^{ij}$ is not determined by the integrated overlap and
we have arbitrarily chosen the normalization of $V$. To find
$N^{ij}$, one uses the unintegrated $Q^\mu$ overlap of equation (2.41)
and one finds the result to be
$$
N^{ij} = -\ln |(z_j-z_i)|.
\eqno(2.47)
$$
One could, from the beginning, have determined $V$ completely
from the $Q^\mu$ overlap, however, it is technically simpler to first
carry out the partial determination given above with the integrated
$P^\mu$ overlap. It is straightforward to evaluate the vertex for any
choice of $(V_i)^{-1}$, or local coordinates.
\par
We have used a somewhat short hand notation in equation (2.46).
By
$\prod_{k=1}^N {}_{(k)}\langle 0|$ we actually mean
$$\prod_{k=1}^N \int d^D q{}_k  \delta\left(\sum_l q^\mu_l\right)
{}_{(k)} \langle 0, q_k| \eqno(2.48)
$$
where $|0,p\rangle$ obeys
$\alpha_n^\mu |0,p\rangle=0,n\ge 1$ and $p^\mu$ is its momentum.
\par
We find that a small change, $z_j \rightarrow\hat z_j= z_j +
\epsilon$, $z_i \rightarrow \hat z_i=z_i$ for $i \neq j$ implies
that $(V_j)^{-1} \hat V_j(z)= z+\hat z_j-z_j=z+\epsilon$ and
$(V_i)^{-1} \hat V_i(z)= z$ for $i \neq j$. The induced
transformation has ${\cal N}_j=\exp (\epsilon L_{-1}^{(j)})$ all
conformal transformations on the other string Hilbert spaces being
the identity operator. As a result equation (2.24) implies that
$$
{\partial V \over \partial z_j} = V L_{-1}^{(j)}. \eqno(2.49)
$$
\par
We can  now determine the measure. Applying the zero norm physical state
$L_{-1}^{(j)} |\Omega\rangle_j$ on the $j$th string Hilbert space and
 arbitrary physical states $|\chi_k\rangle$ to all the other Hilbert spaces
to the integrated vertex
$$
W = \int \prod_{j=1}^{N-3} dz_j f_m V
\eqno(2.50)
$$
we find that
$$
W|\chi\rangle_{(1)} \ldots |\chi\rangle_{(N)} = \int
\prod_{k=1}^{N-3} dz_k {f_m} {\partial V \over \partial z_j}
|\chi\rangle_{(1)} \ldots |\Omega\rangle_{(j)} \ldots
|\chi\rangle_{(N)} \quad ; \quad j = 1, \ldots,N-3 \eqno(2.51)
$$
which implies that ${\partial {f_m} \over \partial z_j} = 0$, or
that ${f_m}$ is a constant. One must also show that the state
$L_{-1}^{(j)}|\Omega\rangle_{(j)}$ decouples for $j = N-2, \ N-1,
\ N$. This is achieved by  using the three  vertex preserving
integrated overlaps of equation (2.30), which follow from taking
$\varphi$ to be the three analytic vector fields that live on the
Riemann sphere, namely $\varphi=1,z,z^2$. This is related to the
${\rm SU}(1,1)$ invariance of the result which is discussed using
these identities in reference [14]. To complete the discussion of
decoupling, one must also show that states of the form $(L_{-2} +
{3\over2} L_{-1}^2)|\Omega'\rangle$ decouple.   This is shown
using
 the vertex  preserving integrated overlaps containing
$L_{-2}^j$ on one leg.
\par
Thus we have completely determined the open string tree level vertex $V$
and the measure
${f_m}$. To find the scattering amplitude for any desired physical
states, one simply saturates $W$ with them.
\par
The derivation of the vertex
for genus greater than one follows a similar path. However, one must
consider integrated overlap identities for functions that are allowed to
shift by constants under the periodicity relations of the surface. In
this case one also finds contributions coming from the boundaries that
represent the Riemann surface. This allows one to take account of the fact
that strictly periodic scalar functions having  only a single pole at a
given point do not exist if the pole is of certain degrees.
\par
We note that in the older literature on the group theoretic approach
(except the more recent [22]),
the overlap equations look a bit
different because a different convention for a conformal operator was
used. In this paper we use the more
common definition given in equation (2.1).
while in the previous literature, $(\xi_i)^{-d}$ was factored
out of the operator, which led to the
transformation rule
$$
{R(\xi_i) \over (\xi_i)^d} \rightarrow {R(\xi_j) \over (\xi_j)^d} \left({d
\xi_j \over d \xi_i} \right)^d.
\eqno(2.52)$$
\par
The same group theoretic approach can be applied to superstrings provided
one makes the appropriate generalisations. We refer the reader to
references [10,19,20] for such discussion.
\par
Although the group theoretic approach has so far only been  used
to construct the scatterings in string theories that are based on
free field constructions, it could be used for string theories in
general and in particular those based on $N=2$ minimal models. The
latter    are associated with dimensional reductions on Calabi-Yau
spaces. The conformal operators are the primary fields that occur
in the minimal models and lead to corresponding overlap equations.
The measure is then determined by the decoupling of zero norm
physical states.

\bigskip
{\bf 3 High Energy Tree Level Arbitrary String Scattering}
\medskip
We consider the limit in which $\alpha^\prime\to \infty$, but the
angle of scattering is fixed. In this section we find the
behaviour of open bosonic strings at the tree level. In this limit
$\alpha_0\to \infty$ and $V\alpha_0^{\mu (i)} \gg V\alpha_n^{ \mu
(i)}$ for $n$ a positive integer, and so we neglect terms of the
latter form. However, we must keep terms of the form
$V\alpha_{-n}^{\mu (i)}$ for $n$ a positive integer as these may
lead to $V\alpha_0^{\mu (j)}$ using the overlap relations. We
could just extract the high energy vertex from equation (2.46),
but it is more instructive to derive the vertex from first
principles in the high energy limit.
\par
We adopt the simple local coordinate systems of equation (2.39).  Taking
$R=\partial x^\mu$, the overlap equation (2.15),  for the functions of
equation (2.43), and in the high energy limit reads
$$
V\left(\alpha_{-n}^{ \mu (i)}+\sum_{j,j\not=i} {\alpha_0^{\mu (j)}\over
(z_j-z_i)^n}\right)=0
\eqno(3.1)$$
As such, the vertex is of the form
$$
V=\prod_{k=1}^N {}_{(k)}\langle 0|\exp\left(-\sum_{i \neq
j}\sum_{n=1}^{\infty} \alpha_n^{\mu(i)}{1\over
n(z_j-z_i)^n}\alpha_{0\mu}^{(j)} +{1 \over 2} \sum_{i \neq j}
\alpha_0^{\mu(i)}\ln|(z_j-z_i)|\alpha_{0\mu}^{(j)}\right)
\eqno(3.2)
$$
The last term does not follow from the integrated
overlap equation (3.1), but is derived from the unintegrated
overlap involving $Q^\mu$. We note that in the high energy limit
we have neglected the term involving $\alpha_n^{  (i)}\cdot
\alpha_m^{  (j)}$ in the general vertex of equation (2.45).
Naively, one may be tempted to also drop the first term in the
vertex in equation (3.2) with respect to the second term, however,
this is required to give the leading terms in the scattering of
strings which are not tachyons.
\par
We now consider the extremisation of the vertex with respect to the
positions of the Koba-Nielsen points $z_i$. Equation (2.49) is valid at
any energy and the vertex will be an extremum if ${\partial V\over
\partial z_i}=0$ and, as a result,
$$
VL_{-1}^{(i)}=0
\eqno(3.3)$$
However, in the high energy limit $L_{-1}^{(i)}=\alpha_{-1}^{
(i)}\cdot \alpha_0^{(i)}$, and so we conclude that
$$
0=V\alpha_{-1}^{ (i)}\cdot \alpha_0^{  (i)}=-
V\sum_{j,j\not=i}{\alpha_0^{  (i)}\cdot \alpha_0^{  (j)}\over (z_j-z_i)}
\eqno(3.4)$$
where to find  the last line we have used equation (3.1).
Hence we conclude that the scattering at high energy is dominated by
momenta that satisfy the condition
$$
\sum_{j,j\not=i}{p_{  i}\cdot p_{  j}\over (z_j-z_i)}=0
\eqno(3.5)$$
\par
Let us examine the consequence of this relations for four particle
scattering. Taking $i=1$ equation (3.5) reads
$$
{p_{  1}\cdot p_{  2}\over (z_1-z_2)}+{p_{  1}\cdot p_{  3}\over
(z_1-z_3)}+{p_{  1}\cdot p_{  4}\over (z_1-z_4)}=0
\eqno(3.6)$$
Using momentum conservation, i.e. $ p_{  4}=-( p_{  1}+ p_{  2}+ p_{
3})$,  the fact that in the high energy limit $ p_{  i}\cdot  p_{i}=0$
and  the usual variables
$$
s=-(p_1+p_2)^2,\ \ t=-(p_1+p_3)^2,\ \ u=-(p_1+p_4)^2 ,
\eqno(3.7)$$
equation (3.6) becomes
$$
{s\over t}= {p_{  1}\cdot p_{  2}\over p_{  1}\cdot p_{  3}}=-
{(z_1-z_2)(z_3-z_4)\over (z_1-z_3)(z_2-z_4)}
\eqno(3.8)$$
where on the right-hand side we recognise the cross ratio. Taking other
values of $i$ we recover the same equation, once we use the relations
between the various cross ratios. Thus we recover the result of references
[2,3] found in the context of tachyon scattering.
\par
Any physical state is of the form
$$
\alpha_{-n_1}^{ \mu_1}\ldots \alpha_{-n_p}^{ \mu_p}|0,p\rangle
\eqno(3.9)$$
where $|0,p\rangle$ satisfies $\alpha_n^\mu |0,p\rangle=0,n\ge 1$
and $\alpha_0^\mu
|0,p\rangle=\sqrt {2\alpha^\prime}p^\mu |0,p\rangle$. Acting with the vertex on
this
state, positioned in the
$i$th string Hilbert space, we may use the overlap relation of  equation
(3.1) to swop any $\alpha_{-n}^{\mu (i)}$ for
$-\sum_{j,j\not=i}{\alpha_{0}^{\mu (j)}\over (z_j-z_i)^n}$.  Once all
the oscillators have been traded in this way one is left with an
expression which is polynomial in momenta times the tachyon amplitude.
\par
For example the scattering of four photons in the high energy limit
becomes
$$V\prod_k\epsilon^{\mu (k)}\alpha_{-1}^{(k)}{}_{
\mu}|0,p_k\rangle_{(k)}=
\prod_i \sum_{j, j\not= i}{1\over (z_j-z_i)}
\epsilon^{\mu (i)}\alpha_{0\mu}^{(j)} \left( V\prod_k
|0,p_k\rangle_{(k)}\right)
\eqno(3.10)$$
where in  the  expression in brackets at the end of this equation we
recognise the tachyon amplitude.
\par
These are the symmetry relations found in reference [5] at high energy.
They are a consequence of using  the overlap relations of the group
theoretic approach involving
$R=\partial x^\mu$  to rewrite the physical states acting on the vertex.
However, the overlap relations are valid at all energies and resulting
identities can be derived relating all scattering amplitudes to those
of the tachyons. At high energy  the identities
simplify considerably.
\par
The closed string proceeds along very similar lines. Equation (3.1) holds
with an analogous equation for $\bar \alpha_n^\mu$. The vertex is given
by equation (3.2) with the addition of very similar terms for the barred
oscillators which involve $\bar z_j$'s. Equations (3.3) to (3.8) carry
over unchanged.  We also conclude that all amplitudes can be
expressed in terms of the tachyon amplitude using the overlap relations,
which are the corner stone of the group theoretic approach,  at any
energy, but these simplify considerably at high energy.


\bigskip
{\bf 4 High Energy Genus One Arbitrary String Scattering}
\medskip
In this section we will find the one loop amplitudes for arbitrary open and
closed
string scattering in the limit of high energy, i.e.
$\alpha^\prime\to \infty$, but with the angle of scattering  fixed. As
at the tree level, in this limit
$\alpha_0\to \infty$ and
$V\alpha_0^{\mu (i)} \gg V\alpha_n^{ \mu (i)}$
for $n$ a positive integer, and so we neglect terms of the latter form.
However, we must keep terms of the form $V\alpha_{-n}^{\mu (i)}$ for $n$ a
positive integer as these may lead to $V\alpha_0^{\mu (j)}$ using
the overlap relations.
We will carry out the calculation using two different representations of the
genus one surface. In the first sub-section we use the Fuchsian
representation involving two identifications of the opposing sides of a
parallelogram and in the next sub-section we represent the genus one
surface by the Riemann sphere with two square root branch cuts.   We also
discuss the relationship between the results found in the two
representations and explain how the differences with the results of
reference [5], which used the latter representation, arise.
\medskip
{\bf 4.1 Genus One Scattering using the Fuchsian Representation}
\medskip
 We could just extract the high energy vertex from the general one loop
vertex calculated in section six, but it is more instructive
to derive the vertex from first principles in the high energy limit. As
such, this section is largely self contained.
\par
 At genus one the string sweeps out a  surface that is a torus with marked
points. A torus can be thought of as the complex plane ${\bf C}$  subject
to the equivalence relation $z\sim z+2nw+2mw^\prime$
for $n,m\in {\bf Z}$ for all points $z\in
{\bf C}$. The torus is then represented by a
 parallelogram whose   corners we may choose to be  at the points
$0,2w,2w+2w^\prime$ and $2w^\prime$ as one moves around the
parallelogram  in an  anti-clockwise movement. For the closed string we
choose $2w=1$, while for the doubled surface for the open string we write
$2w=-\ln u$, and take
$w^\prime=i\pi$.   The torus is formed by
identifying its opposing sides of the parallelogram, which is enforced by
the equivalence relation. Thus the torus  has the periods $2w$ and
$2w^\prime$ with modulus parameter defined to be $\tau\equiv
{w^\prime\over w}$.  The region inside the parallelogram is defined to be
the fundamental domain and its points are in one to one correspondence
with those of the genus one surface.  We use the conventions of Bateman
[24] and will refer to this representation of the genus one surface as
the Fuchsian representation.  We will adopt the simple local coordinates
$\xi_i=z-z_i$, of  equation (2.39), in the neighbourhood of the Koba-Nielsen
points $z_i$.
\par
We begin by considering the scattering of four closed strings.
We now consider the  overlap equations
(2.5) involving the conformal field
$\partial x^\mu$.
Using these and the overlap relations
$$
V\partial x^\mu(z)=V\partial x^\mu(z+2w),\ \ \
V\partial x^\mu(z)=V\partial x^\mu(z+2w^\prime),
\eqno(4.1.1)$$
which are those of equation (2.8) for the conformal operator $\partial
x^\mu$, we can derive the integrated overlap  equation (2.15), namely
$$\sum_j \oint d\xi_j V\varphi\partial x^{\mu(j)}(\xi_j)=0
\eqno(4.1.2)$$
where  $\varphi$ is  a scalar which
satisfies the periodicity conditions
$$
\varphi(z)=\varphi(z+2w),\ \ \varphi(z)=\varphi(z+2w^\prime)
\eqno(4.1.3)$$
and is analytic except for possible  poles at the Koba-Nielsen
points. The derivation is as explained in section two, the cancellation
of contributions from opposite sides of the parallelogram being obvious
as $\varphi$ and $VR$ are periodic.
\par
To derive the vertex it is desirable to have a periodic function
with a pole at only one Koba-Nielsen point. The Weierstrass function
${\cal P}(z)$ is the unique function which  has a pole at order two of
the form
${1\over z^2}$ at $z=0$, and points related by the periods of the torus,
but   is otherwise analytic. Its derivatives  ${\cal P}^{(m)}(z)\equiv
{d^m {\cal P}(z)\over dz^m}$ have poles of  all higher orders. There is
no such function with a pole of order one. This is obvious,   integrating
such a possible periodic function around the boundary of the
parallelogram  and using equation (4.1.3) we find zero. However,  this is
also the residue to the pole which  must therefore vanish.  We will discuss
below how to deal with this case. Taking $\varphi={(-1)^m\over
(m+2)!}{\cal P}^{(m)}(z-z_i)$, the overlap equation (4.1.2), in the high
energy limit, reads for $n \ge 2$
$$ V\left({1 \over n} \alpha_{-n}^{ \mu (i)} +
{(-1)^n \over n!} \left. {d^{n-2} \over dt^{n-2}} ({\cal P}(t) -
t^{-2}) \right|_{t=0}\alpha_0^{\mu (i)} +\sum_{j,j\not=i}
{(-1)^n\over n!}{\cal P}^{(n-2)}(z_j-z_i)\alpha_0^{\mu (j)}
\right)=0. \eqno(4.1.4)
$$
Note that in the high energy limit we
keep only negative moded oscillators and the momenta acting on the
vertex. As such, the only information required from $\varphi$ is
the degree of the pole at the one Koba-Nielsen point and its
values at the other Koba-Nielsen points.
\par
We now restrict our attention to the scattering of four closed
strings. It has been shown [2,3] that the genus one closed string
amplitude at high energy is dominated when the Koba-Nielsen points
sit at the half periods of the torus. We now consider only such
scattering, but we will show below that the amplitude is
extremised in this configuration. We therefore take the
Koba-Nielsen points to be at
$$
z_1=w,\ z_2=w+w^\prime,\ z_3=w^\prime,\ z_4=0
\eqno(4.1.5)$$
Since the Koba-Nielsen points are on the edges of the fundamental
regions we should specify how we draw the integration contour. We will
adopt our contour of integration used in the integrated overlap
equations  to start at $-w^\prime$, not including this point inside the
contour as it goes anti-clockwise around,  move to the point
$-w^\prime+2w$ not including this point, then move to $w^\prime+2w$
not including this point, then to
$w^\prime$  including this point and back to the beginning. As we do
so  the contour is deformed so that the Koba Nielsen points within the
contour are at $0$, $w$, $w+w'$ and $w'$ with all their images
excluded from the contour. We have drawn this in figure 1.

\input epsf
\medskip
\centerline{\epsfbox{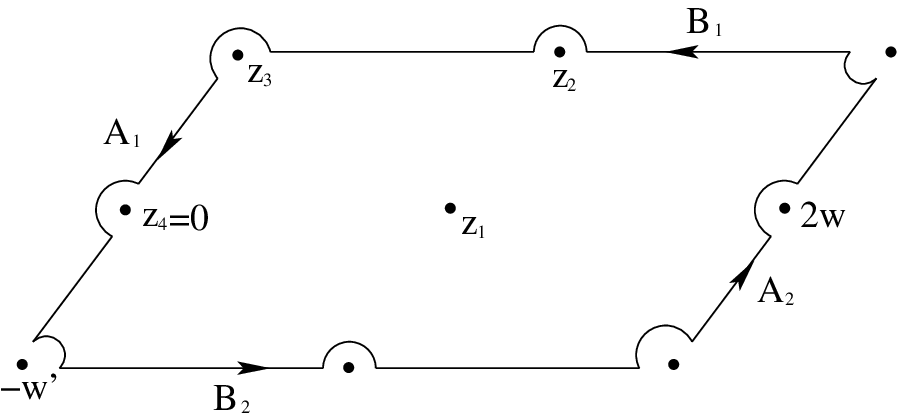}}
\medskip
\centerline{{\bf Figure 1:} The contour}
\medskip

\par
Now it is known [24] that
${\cal P}^{(2m+1)}$, for  integer $m=1,2,3,\dots $, is equal to ${\cal
P}^{(1)}$ times a polynomial of degree $m$ in ${\cal P}$. However,
${\cal P}^{(1)}$ vanishes at the half periods, ${\cal
P}^{(1)}(z_i-z_j)=0$ and so we find that
$$
V\alpha_{-(2m+1)}^{ \mu (i)}=0,\ \ m=1,2,3,\ldots
\eqno(4.1.6)$$
We note that ${\cal P}^{(2m-2)}$, for positive integer $m$, is equal to
 a polynomial of degree $m$ in ${\cal P}$ which does not in general
vanish at the half periods.
\par
As we noted above, there does not exist a periodic function that has only
a single pole of order one.  The way around this difficulty is to consider
functions which are not periodic, but shift by constants under the change
of periods, namely a  function such that
$$
\varphi(z+2w)=\varphi(z)+c_1,\ \
\varphi(z+2w^\prime)=\varphi(z)+c_2
\eqno(4.1.7)$$
We now consider the contour integral of $V\partial x^\mu\varphi$ around
the boundary of the torus. The contributions from opposing sides no
longer cancel and one finds instead of equation (4.1.2)  that
$$
\sum_j \oint d\xi_j V\varphi\partial
x^{\mu(j)}(\xi_j) = c_2\int_{B_1}V\partial x^{\mu}dz
+c_1\int_{A_2}V\partial x^{\mu}dz
\eqno(4.1.8)$$
On the right hand side $V\partial x^{\mu}dz$ stands for $V\partial
x^{\mu(k)}(\xi_k) d\xi_k$ which is independent of which string Hilbert
space one uses by virtue of equation (2.5) using the operator $\partial
x^{\mu}$.  The vertex  should not depend on  which function $\varphi$ one
takes as long as it obeys equation (4.1.7). Functions of this type,  with
a  pole of order one at the same place, differ by a constant times $z$.
Taking
$\varphi=z$, which of course is  a function which shifts by periods
of the torus, in equation (4.1.8) we find that provided
$$
 w^\prime\int_{B_1}V\partial x^{\mu} dz
+w\int_{A_2}V\partial x^{\mu}dz={1\over 2}V\sum_j z_j
\alpha_0^{\mu (j)}
\eqno(4.1.9)$$
we will indeed find the same result no matter what function $\varphi$ we
take. In this last equation we have no $\alpha_1^\mu$ terms as we are
in  the  high energy limit.   As we will show in section six,  modular
invariance allows one to conclude that the solution to equation (4.1.9) is
given by
$$
\int_{B_1}V\partial x^{\mu}dz ={1\over 2}V(
\alpha_0^{\mu (2)}+\alpha_0^{\mu (3)})
\eqno(4.1.10)$$
\par
The function
$$
\varphi_1(z)= \zeta(z)-{\eta\over w}z ,
\eqno(4.1.11)$$
 obeys equation (4.1.7) with $c_1=0$ and $c_2=-{i\pi\over w}$. In this
equation $\zeta(z)$ is the zeta function which is defined to obey
$-{d\zeta(z)\over dz}={\cal P}(z)$ and has a first order pole with residue
one.  It is not a periodic function, but obeys
$$
\zeta(z+2w)=\zeta(z)+2\eta,\ \ \ \zeta(z+2w^\prime)=\zeta(z)+2\eta^\prime
\eqno(4.1.12)$$
where $\eta$ and $\eta^\prime$ are independent of $z$.
Substituting $\varphi_1$  into
equation (4.1.8)   and using equation (4.1.10) we find that
$$
V\left(\alpha_{-1}^{ \mu (i)}+\sum_{j,j\not=i} \alpha_0^{\mu (j)}
{\varphi_1}(z_j-z_i)\right)=-{i\pi\over2 w}V(
\alpha_0^{\mu (2)}+\alpha_0^{\mu (3)})
\eqno(4.1.13)$$
Evaluating $\varphi_1(z_j-z_i)$ we find that this equation becomes
 $$
V\alpha_{-1}^{ \mu (i)}=0 ,
\eqno(4.1.14)$$
which, together with equation (4.1.6), implies that all odd oscillators
acting on the vertex vanish. The story for the overlap equations
involving $\bar \alpha_n^\mu$ is very similar.
\par
Using equations (4.1.4), (4.1.6)  and (4.1.14) we find that the
high energy vertex is given by
$$
V=\prod_{k=1}^N {}_{(k)}\langle 0|\exp\left(-\sum_{i\neq j}
\sum_{n=1}^{\infty}\left(\alpha_{2n+2}^{ \mu (i)}{{\cal P}
^{(2n)}(z_j-z_i)\over (2n+2)!}\alpha_{0\mu}^{(j)} +\bar
\alpha_{2n+2}^{ \mu (i)}{\bar{\cal P} ^{(2n)}(\bar z_j-\bar
z_i)\over (2n+2)!}\alpha_{0\mu}^{(j)}\right) \right.
$$
$$
-\sum_i\sum_{n=1}^{\infty}\left(\alpha_{2n+2}^{\mu(i)} \left.
{d^{2n} \over dt^{2n}} {({\cal P}(t) - t^{-2}) \over (2n+2)!}
\right|_{t=0} \alpha_{0\mu}^{(i)} + \bar{\alpha}_{2n+2}^{\mu(i)}
\left. {d^{2n} \over dt^{2n}} {(\bar{\cal P}(t) - t^{-2}) \over
(2n+2)!} \right|_{t=0} \bar{\alpha}_{0\mu}^{(i)} \right)
$$
$$\left. +\sum_{i \neq j}
\alpha_0^{ \mu (i)}\ln |\sigma(z_j-z_i)|\alpha_{0\mu}^{(j)}\right)
\eqno(4.1.15)
$$
The last term does not follow from the integrated overlap equations,
but is derived from the unintegrated overlap involving $x^\mu$.
\par
The high energy limit is given by extremising the vertex with respect to
the Koba-Nielsen points and the moduli.
The vertex will be an extremum with respect to the Koba-Nielsen
coordinates if
${\partial V\over
\partial z_i}=0$. For the simple local coordinates which
we are using equation (2.49), which is also valid at any genus and
energy,  implies that
$$
VL_{-1}^{(i)}=0
\eqno(4.1.16)$$
In the high energy limit, $L_{-1}^{(i)}=\alpha_{-1}^{
(i)}\cdot \alpha_0^{  (i)}$ and using equation (4.1.14) this is
automatically satisfied if the Koba-Nielsen points are at the half
periods.
\par
We now find the consequences of the vertex being at an extremum of the
modulus of the surface. To this end we need to find which conformal
transformation changes the moduli i.e. $w^\prime \to w^\prime +\epsilon$
as we have taken $2w=1$. As such we need an $f$ in equation (2.25) such
that
$$f(z)=f(z+2w),\ \ \ f(z)=2+f(z+2w^\prime)
\eqno(4.1.17)$$
Such a function is given by
$$
f(z)={2w\over i\pi}\left(
 \zeta(z)-{\eta\over w}z\right)
\eqno(4.1.18)
$$
Substituting this function with argument $z-z_i$ into equation (2.27) we
find in the high energy limit the equation
$$
{i\pi\over 2 w}{\partial V\over \partial w^\prime}=VL_{-2}^{(i)}
\eqno(4.1.19)$$
In deriving this equation we have used equation (4.1.16) and the fact that
in the high energy limit $L_n^{(j)}=0, \ n\ge0$ and $c_f$ vanishes
as there is no $\alpha_1\cdot \alpha_1$ term in the vertex. The high
energy vertex is an extremum and so
$$VL_{-2}^{(i)}=0
\eqno(4.1.20)$$
This equation and equation (4.1.16) ensure that any zero norm physical
states decouple.
Using equation (4.1.14), equation (4.1.20) implies in the high energy
limit that
$$
V\alpha_{-2}^{(i)}\cdot \alpha_0^{(i)}=0
\eqno(4.1.21)$$
Using equation (4.1.4), we can trade in $\alpha_{-2}^{(i)}$  to find that
$$
V\sum_{j,j\not= i} {\cal P}(z_j-z_i)\alpha_{0}^{(j)}\cdot \alpha_0^{(i)}=0
\eqno(4.1.22)$$
Introducing ${\cal P}(z_k)=\varsigma_k$ the above equation becomes
$$
p_1\cdot p_2 \varsigma_3+p_1\cdot p_3 \varsigma_2+p_1\cdot p_4
\varsigma_1=0
\eqno(4.1.23)$$
and so
$$
{s\over t}= {p_{  1}\cdot p_{  2}\over p_{  1}\cdot p_{  3}}=
-{(\varsigma_1-\varsigma_2)\over (\varsigma_1-\varsigma_3)}
\eqno(4.1.24)$$
where $ s=-(p_1+p_2)^2,\ \ t=-(p_1+p_3)^2,\ \ u=-(p_1+p_4)^2$. As we
explain in appendix A,  $\varsigma_k$ are the images of
$z_k$ when the torus is mapped to the representation of the genus one
surface that consists of the Riemann sphere with two branch cuts,
denoted $\hat C_2$, and the points $\varsigma_k, k=1,2,3, 4$ are the
positions of the branch cuts. We note that    $\varsigma_4=\infty$ and so
the quantity that appears on the right-hand side is a cross ratio.
 Thus we recover the result of references
[2,3] found in the context of tachyon scattering.
\par
We can also express the scattering of any physical states in terms of
momenta multiplied by the tachyon amplitude. The discussion is much
like that for the tree level given in section three. Any physical state is
of the form
$$
\alpha_{-m_1}^{ \mu_1}\ldots \alpha_{-m_p}^{ \mu_p}
\bar{\alpha}_{-n_1}^{ \nu_1}\ldots \bar{\alpha}_{-n_q}^{ \nu_q}|0,p\rangle.
\eqno(4.1.25)$$
If the expression involves any odd moded oscillators then
the amplitude will vanish using equations (4.1.6) and (4.1.14) in the extreme
high energy limit. Using the overlaps relation of equation (4.1.4), the even
moded oscillators can be traded
for momenta.
Once all
the oscillators have been traded in this way one is left with an
expression which is polynomial in momenta times the tachyon amplitude.
However, as at tree level, the overlap relations are valid at all
energies and resulting  identities can be derived relating all scattering
amplitudes to those of the tachyons.
\par
Although the results found above for tachyon scattering agree with those
found in references [2-5] those for other string states differ
considerably. In particular in reference [5] the scattering of states
containing odd moded oscillators does not vanish, but we see that there
are no odd moded oscillators in the vertex and so these amplitudes
vanish in the extreme high energy limit. The simplest example of this is
the four graviton scattering which has a tree level contribution, but no one
loop contribution.
\par
For the open string we take $w^\prime=i\pi$ and $w=-\ln u$ where $u$
is a parameter which appears in the sewing of vertices to find the one
loop amplitude. The planar one loop open string amplitude does not
have an extremum at high energy, but the non-planar amplitude has with
the Koba-Nielsen points at $0$, $w$, $w+w^\prime$ and $w^\prime$, that
is at the half periods of the torus. The calculation for the open  string
proceeds along very similar lines to the closed string with almost all
the equations being as above once one deletes any terms containing
$\bar \alpha_n^\mu$'s. For example,  equations (4.1.1-4.1.15) hold.
 The vertex is given by equation (4.1.15) once one deletes  the term
containing $\bar \alpha_n^\mu$'s. Equations (4.1.16-24) also hold with
almost identical conclusions.

\medskip
{\bf 4.2  Genus One Scattering using the Cut Sphere  Representation}
\medskip
We now repeat the calculation for the scattering of closed strings
at genus one, but now representing the surface by the Riemann
sphere with two square root branch cuts, denoted $\hat C_2$. The
positions of the cuts are chosen to be between $\varsigma_1$ and
$\varsigma_2$, and $\varsigma_3$  and $\varsigma_4=\infty$.  The
Fuchsian representation of the genus one surface, considered in
the previous section, with coordinate $z$, and that of the Riemann
sphere with two branch cuts and coordinate $\varsigma$ considered
here are related by
$${\cal P}(z)=\varsigma, \ \ \ z=\int_\infty^\varsigma
{d\varsigma^\prime\over Q(\varsigma^\prime)}
\eqno(4.2.1)$$
where
$Q=-2\sqrt{(\varsigma-\varsigma_1)(\varsigma-\varsigma_2)
(\varsigma-\varsigma_3)}={\cal P}^{(1)}(z)$. A more detailed discussion
of this mapping will be given in appendix A.
\par
The high energy scattering is dominated by amplitudes for which the
Koba-Nielsen points are at the half periods as in equation (4.1.5). Under
the map to $\hat C_2$ these points get mapped to ${\cal
P}(z_i)=\varsigma_i, i=1,2,3,4$, and are the beginning and end points of the
branch cuts. They also sit on the real axis such that
$\varsigma_1> \varsigma_2 > \varsigma_3$.
\par
The representation of the surface one uses does not by itself affect the
form of the vertex in terms of oscillators as the overlap functions
$\varphi$ used in one representation can be mapped using equation (4.2.1)
to the other representation. However, the form of the vertex does depend
on the local coordinates used. For  string scattering using the
Fuchsian representation we used the coordinate $z$ of the complex plane
and the simple local coordinates
$\xi_i=z-z_i$.  Using equation (4.2.1), we find around the point $z_i$
that
$$
z-z_i=\int_{\varsigma_i}^{\varsigma} {d\varsigma^\prime\over
Q(\varsigma^\prime)}= \sum _{n=0}^\infty a_{2n}^{(i)}
(\varrho_i)^{2n+1}. \eqno(4.2.2)
$$
where
$$
\varrho_i=\sqrt {(\varsigma-\varsigma_i)},\ i=1,2,3,\ \
\varrho_4={1\over \sqrt {\varsigma}}
\eqno(4.2.3)$$
For $i=1$, the lowest terms in this expansion are given by
$$
z-z_1=-{\varrho_1\over
\sqrt{(\varsigma_1-\varsigma_2)(\varsigma_1-\varsigma_3)}}
+{\varsigma_1\over
2[(\varsigma_1-\varsigma_2)(\varsigma_1-\varsigma_3)]^{3\over
2}}(\varrho_1)^{3}+\dots ,
\eqno(4.2.4)$$
Equation (4.2.2) implies  that $\varrho_i$ are good coordinates on $\hat
C_2$ in the neighbourhood of the Koba-Nielsen points $\varsigma_i$.
Indeed,  they are related by an analytic transformation to the
coordinates $\xi_i$. We note that
$\varsigma-\varsigma_i$ are not good local coordinates as they fail to
take account of the branch cuts. Indeed, they  do not give a one to
one map between the surface and a region of the complex plane as good
coordinates are required to  do. We have emphasised this point as it
lies at the heart of the different   results found in  this paper
compared to  those of reference [5].
\par
The two local coordinates $\xi_i$ and $\varrho_i$ are   good local
coordinates on the Fuchsian and   $\hat C_2$ representations of the
Riemann surface respectively. However, as they are related by an analytic
transformation, we can use equation (4.2.2) to define a good set of
$\xi_i$ local coordinates on  $\hat C_2$ and, conversely, a good set of
local coordinates $\varrho_i$ on the  Fuchsian representation. Of course,
the final scattering amplitude is independent of the local coordinates
used,   but the vertex and measure  will be different. The overlap
functions
$\varphi$ used for the Fuchsian representation can be mapped over to
$\hat C_2$ to yield overlap identities on $\hat C_2$. For example, if we
work on
$\hat C_2$, but use the coordinates $\xi_i$ to compute the vertex it will
be exactly the same as the vertex computed using the Fuchsian
representation with coordinates $\xi_i$ simply because the overlap
functions have the same functional dependence on a given set of local
coordinates regardless of which representation of the surface is used.
When working on
$\hat C_2$ it is more natural to use the local coordinates  of equation
(4.2.3) and in this section we will compute the vertex in this coordinate
system.
We note that because the coordinate transformation of equation (4.2.2)
contains only odd powers of $\varrho_i$ the conformal map which relates
the vertex $V$, computed with  $\xi_i$ local coordinates, to the
vertex, denoted $V_\varrho$, computed with  $\varrho_i$ local coordinates,
is given by
$$
V\prod_i {\cal N}_i^{(i)}=V_\varrho,\ \ {\rm where }\ \
{\cal N}_i^{(i)}=\exp\left(\sum_{n\ge 0} c_{2n}^{(i)} L_{2n}^{(i)}\right)
\eqno(4.2.5)$$
\par
Clearly, the functions
$$
\varphi_c={1\over (\varsigma-\varsigma_i)^n}={1\over
(\varrho_i)^{2n}}
\eqno(4.2.6)$$
are well defined functions on $\hat C_2$, even if they   are
insensitive to the branch cuts. These lead to the overlap identities
$$
V_\varrho\left(\alpha_{-2n}^{(i)}+\sum_{j,j \not= i} {1\over
(\varsigma_j-\varsigma_i)^n}\alpha_0^{(j)}\right)=0 \eqno(4.2.7)
$$
Other
functions which are well defined on $\hat C_2$ are given by
$$
\varphi_c^{2n-1}\equiv{Q(\varsigma)\over
(\varsigma-\varsigma_i)^{n}}= {d_{(2n-1)}\over
(\varrho_i)^{2n-1}}+\ldots +{d_1\over \varrho_i}+{\rm analytic} ,\
\ n\ge 2 \eqno(4.2.8)$$ where $d_i$ are constants. The value $n=1$
is excluded as that function has a pole at infinity. In fact,
equations (4.2.6) and (4.2.8) are all the possible functions
defined on $\hat C_2$ with a single pole at one of the
Koba-Nielsen points. They do indeed correspond to linear
combinations of the overlap functions used in the previous
section, as can be seen by using the relation [24]
$${\cal P}(z-z_i)=\varsigma_i+{(\varsigma_i-\varsigma_{i+1})
(\varsigma_i-\varsigma_{i+2})\over
\varsigma-\varsigma_i}
\eqno(4.2.9)$$
where $\varsigma_{i+3}=\varsigma_{i}$. Indeed, ${\cal P}^{(2m-2)}(z-z_1)$
is a polynomial in  ${\cal P}^{(2m)}(z-z_1)$ and so in ${1\over
(\varsigma-\varsigma_i)^m}$, while
$$
{\cal P}^{(1)}(z-z_1)=-{d\varsigma\over dz} {(\varsigma_i-\varsigma_{i+1})
(\varsigma_i-\varsigma_{i+2})\over
(\varsigma-\varsigma_i)^2} = -{(\varsigma_i-\varsigma_{i+1})
(\varsigma_i-\varsigma_{i+2})Q(\varsigma)\over
(\varsigma-\varsigma_i)^2}.
\eqno(4.2.10)$$
Higher order odd derivatives are found using
${\cal P}^{(2m+1)}(z-z_1)$ is a polynomial of degree m  in  ${\cal
P}(z-z_1)$ times ${\cal P}^{(1)}$ and so all odd derivatives give
superpositions of the functions in equation (4.2.8).
\par
To find an overlap equation involving $\alpha_{-1}^{\mu (i)}$, we require
a function with a single first order pole. Such functions do not exist
and so,  as for the Fuchsian representation, we can instead  use the function
$\varphi_1$ of equation (4.1.11) expressed in terms of $\varrho_i$
coordinates. The integral  of equation (4.1.10) over $B_1$ takes the same
values since  the vertices $V$ and $V_\varrho$ are related as in equation
(4.2.5) and in the high energy limit we keep  only terms involving
momentum acting on the vertex. Evaluating the overlap we find that
$V_\varrho\alpha_{-1}^{
\mu (i)}=0 $. This result follows more simply by carrying out the
conformal transformation of equation (4.2.5) using equation (4.1.14).
Using the overlap relations which arise from the functions of equation
(4.2.8) we conclude that
$$
V_\varrho\alpha_{-(2n+1)}^{ \mu (i)}=0,\ \ n\ge 0
\eqno(4.2.11)$$
\par
This equation and equation (4.2.7) imply that the vertex, using the
coordinates $\varrho_i$, is given by
$$
V_\varrho= \prod_{k=1}^N {}_{(k)}\langle 0|\exp\left(-\sum_{i\neq
j}\sum_{n=1}^{\infty}\left( \alpha_{2n}^{ \mu (i)}{1\over
2n(\varsigma_j-\varsigma_i)^n}\alpha_{0\mu}^{(j)} +\bar
\alpha_{2n}^{ \mu (i)}{1\over
2n(\bar\varsigma_j-\bar\varsigma_i)^n}\alpha_{0\mu}^{(j)}\right)
\right.
$$
$$ \left. +\sum_{i \neq j}
\alpha_0^{ \mu (i)}\ln \sqrt{|( \varsigma_j-
\varsigma_i)|}\alpha_{0\mu}^{(j)}\right) \eqno(4.2.12)
$$
The last term in the exponential comes from the overlap identity
$V_\varrho Q^{\mu (i)}(\varrho_i)=V_\varrho Q^{\mu (j)}(\varrho_j)$ and is
in agreement with that found in references [2-5]. However, the rest of the
vertex disagrees with that found in reference [5].
\par
We observe that the vertex has a spectacularly simple form, and is just
like that found at tree level in equation (3.2) except  that all the odd
moded oscillators are absent and we make the replacement
$(z_j-z_i)\to \sqrt{(\varsigma_j-\varsigma_i)}$.
\par
The extremum conditions on the vertex are much like that for the
$\xi_i$ coordinates.
 Clearly, $V_\varrho L_{-1}^{(i)}=0$ and the reader will enjoy recovering
equation (4.1.23) from the condition $V_\varrho L_{-2}^{(i)}=0$ using equation
(4.2.7) for $n=1$. We note that this last calculation is much like the
analogous calculation  at tree level.
\par
The results for the open string essentially follow the same path as
explained in the previous section.
\par
We now comment on the relation of the results  contained in this paper
with those found  in  references [2-5]. The four tachyon scattering at
high energy found here agrees precisely with that found in references
[2-4]. While almost all  of references [2-4] concentrate on tachyon
scattering,  reference [5] gives genus independent relations between an
arbitrary string scattering amplitude and that  for the same number of
tachyons. The results for arbitrary string scattering found here disagree
with those found in reference [5]. Indeed, some of the amplitudes, such
as that for graviton scattering, discussed in [5] are found to vanish at
one loop, but are non-zero at tree level, and as a result we do not find
the genus independent relations found in reference [5].
\par
Reference [5] uses the sum
over world histories approach within the context of the representation of
the string world-surface by the Riemann sphere with two  branch cuts of
degree
$g$, where $g$ is the genus of the surface. By degree $g$ we mean that the
two branch cuts are of the form
$(z)^{1\over g+1}$, and so lead to a surface with
$g+1$ different sheets.   For genus one this  is the representation of
the torus with two square root branch cuts used  above. While we agree
with reference [5] that the part of the vertex  which depends only on the
momenta can be expressed in terms of the positions of the branch cuts
written in terms of the coordinate
$\varsigma$ of the  Riemann sphere with  two branch cuts, the source of
our disagreement stems from the use of such a coordinate to describe the
behaviour of the other  physical states at high energy. It is important to
remember that the high energy behaviour is dominated when the Koba-Nielsen
points are at  the beginning and end of the branch cuts.
 As spelt out below  equation (4.2.4)  $\varsigma-\varsigma_i$ are
not a good set of local  coordinates to use when considering arbitrary
string scattering from points that are at the beginning and end of the
branch cuts, as they do not take account of the branch cuts there. As a
result, they  can not be used to formulate the scattering of arbitrary
string states at high energy as was done in reference [5].

\bigskip
{\bf {5. The Purely Momentum Dependence of  String Vertices.}}
\medskip
In this section, we will compute the purely  momentum dependence
of the closed and open string vertices (i.e the part
of the vertex involving only
$\alpha_0^\mu$) in terms of the coefficients that occur
in  the rest of the vertex.  We do this in such a way that it is valid for
the vertex at any genus with any number of strings scattering.
The purely momentum piece of the vertex for planar open string scattering
at any genus  was found in [13,14].
\par
The part of the vertex involving non-zero modes (i.e. $\alpha_n^\mu, \
n\ge 1$)  can be deduced using the integrated overlap equations. This has
been carried out [13,14] for all genus vertices for planar open string
scattering and the computation for the closed string vertices is similar.
In section six we will do the computation for the closed string  and
non-planar open string genus one vertices. We can
 write the vertex in the form
$$
V^{\rm closed} = \prod_{k=1}^N {}_{(k)}\langle 0| \exp \left(-
\sum_{i,j} \left\{ {1\over2}\sum_{n,m=1}^\infty \alpha_n^{\mu {(i)}}
N_{nm}^{ij}
\alpha_{m \mu}^{(j)} +  \sum_{n=1}^\infty \alpha_n^{\mu {(i)}} N_n^{ij}
\alpha_{0 \mu}^{(j)} \right. \right.
$$
$$ \left. \left.  + (N_{nm}^{ij}\to \bar N_{nm}^{ij},\ N_n^{ij}\to \bar
N_n^{ij},\
\alpha_n^{\mu (k)}\to \bar \alpha_n^{\mu (k)})
\matrix{{}\cr {}}\right\} \matrix{{}\cr {} \cr {}}\right) V_{00}^{\rm closed}.
\eqno{(5.1)}
$$
Now we want to find the zero
mode piece, which we write
$$
V_{00}^{\rm closed} = \exp \left(-{1\over2} \sum_{i,j}\alpha_0^{\mu (i)}
N^{ij} \alpha_{0 \mu}^{(j)} \right).
\eqno{(5.2)}
$$
The purely momentum piece can not be calculated from the integrated
overlap equations as they involve no operator ( i.e $q^\mu$) that is
sensitive to this term. To calculate the
$N^{ij}$ we need to use the overlap with
$x^\mu$. The mode expansion of the embedding field, denoted $x^\mu$,  can be
written as
$$
x^\mu(z,\bar{z}) = q^\mu - i {\alpha' \over 2} p^\mu \ln |z|^2 + i
\sqrt{\alpha' \over 2} \sum_{n \neq 0} {1 \over n} \left( \alpha_n^\mu
z^{-n} + \bar{\alpha}_n^\mu \bar{z}^{-n} \right).
\eqno{(5.3)}
$$
We have $\alpha_0^\mu = \bar{\alpha}_0^\mu = \sqrt{\alpha' \over 2}
p^\mu$ and, from the canonical commutation relation $[p^\mu, q^\nu] =
-i \eta^{\mu\nu}$, we can write
$$
q^\mu = -i \sqrt{\alpha' \over 2} {\overleftarrow{\partial} \over
\partial \alpha_{0 \mu}},
\eqno{(5.4)}
$$
and we can therefore factor out the universal $i \sqrt{\alpha' \over
2}$ factor and write
$$
x^\mu(z,\bar{z}) = i \sqrt{\alpha' \over 2}
\left(-{\overleftarrow{\partial} \over \partial \alpha_{0 \mu}} -
\alpha_0^\mu \ln |z|^2 + \sum_{n \neq 0} {1 \over n} \alpha_n^\mu z^{-n} +
\sum_{n \neq 0} {1 \over n} \bar{\alpha}_n^\mu \bar{z}^{-n} \right).
\eqno{(5.5)}
$$

\par
The unintegrated overlap for $x^\mu$ is given by
$$
V x^{\mu (i)}(\xi_i, \bar{\xi}_i) = V x^{\mu (j)}(\xi_{j}, \bar{\xi}_j),
\eqno{(5.6)}
$$
where $V$ is given by (5.1) and (5.2), and we choose the local
coordinates $\xi_i = z-z_i$ where $z$ is the coordinate on the Fuchsian
or Schottky  representation of the Riemann surface. For simplicity, we
will not write anymore the spacetime indices as it is obvious how they
appear. The overlap equation (5.6), using equation (5.5)  and the
form of the vertex of equation (5.1), becomes
$$
V \left\{\sum_kN^{ik} \alpha_0^{(k)} + \sum_{n=1}^\infty \sum_k \left(
\alpha_n^{(k)} N_n^{ki} +
\bar{\alpha}_n^{(k)} \bar{N}_n^{ki} \right) - \alpha_0^{(i)} \ln |\xi_i|^2
\right.
$$
$$
\left. + \sum_{n,m=1}^\infty \sum_k \left(N_{nm}^{ik} (\xi_i)^n
\alpha_m^{(k)} +
\bar{N}_{nm}^{ik} (\bar{\xi}_i)^n \bar{\alpha}_m^{(k)} \right) +
\sum_{n=1}^\infty \sum_k
\left( N_n^{ik} \alpha_0^{(k)} (\xi_i)^n + \bar{N}_n^{ik}
\bar{\alpha}_0^{(k)} (\bar{\xi}_i)^n \right) \right.
$$
$$
\left. + \sum_{n=1}^\infty \left( {1 \over n} \alpha_n^{(i)} (\xi_i)^{-n}
+ {1 \over n} \bar{\alpha}_n^{(i)} (\bar{\xi}_i)^{-n} \right) \right\} = V
\left\{ i \leftrightarrow j \right\},
\eqno{(5.7)}
$$
where the right-hand side is obtained from the left-hand side by
changing all $i$'s by $j$'s.
\par
Let us start by considering only the
$\alpha_0$ terms, and assume that $N_n^{ik}$ has the form
$$
N_n^{ik} = - {(-1)^n \over n!} \left. {d^n \over dt^n}
\left( \ln \psi(z_k-z_i+t) \right) \right|_{t=0} \qquad k \neq i
$$
$$
N_n^{ii} = - {(-1)^n \over n!} \left. {d^n \over dt^n} \left( \ln
\psi(t) - \ln t \right) \right|_{t=0},
\eqno{(5.8)}
$$
and similarly for $\bar{N}_n^{ik}$, but with $\psi(z)$ replaced by
$\bar{\psi}(\bar{z})$, and that
 $(\ln \psi(z) -\ln z)$ is analytic in
a neighborhood of $z=0$. These relations are a consequence of the
integrated overlap identities that determine the non-zero mode part of
the vertex. Indeed,
$\ \ \ $
$- {(-1)^n \over n!} \left. {d^n \over dt^n} \left(
\ln
\psi(z-z_i+t) - \ln (z-z_i+t) \right) \right|_{t=0}$ for $n\ge 1$ are
closely related to the functions used in the overlap relations for the
operators  $\partial x^\mu$.  The assumptions (5.8) allow
us to write the second line of (5.7) in a simple manner
$$
\sum_{n=1}^\infty N_n^{ik} (\xi_i)^n = - \sum_{n=1}^\infty
{(-\xi_i)^n \over n!} \left. {d^n \over dt^n}
\left( \ln \psi(z_k-z_i+t) \right) \right|_{t=0} =
\ln \left({\psi(z_k-z_i) \over \psi(z_k-z)}\right) \qquad k \neq i
$$
$$
\sum_{n=1}^\infty \xi_i^n N_n^{ii} = - \ln \left({\psi(-\xi_i)
\over -\xi_i} \right)
+ \lim_{t \rightarrow 0} \ln \left({\psi(t) \over t} \right),
\eqno{(5.9)}
$$
and similarly for the complex conjugates. The overlap (5.7) thus
becomes, for the $\alpha_0$ terms
$$
V \left\{ \sum_kN^{ik} \alpha_0^{(k)} + \sum_{k, k \neq i} \ln
\left|{\psi(z_k-z_i) \over \psi(z_k-z)} \right|^2 \alpha_0^{(k)}
\right.
$$
$$
\left. + \left( - \ln |\xi_i|^2 - \ln \left| {\psi(-\xi_i)
\over - \xi_i} \right|^2 + \lim_{t \rightarrow 0}
\ln \left| {\psi(t) \over t} \right|^2 \right) \alpha_0^{(i)} \matrix{{}\cr {}
\cr {}}\right\}
= V \left\{ i \leftrightarrow j \right\}.
\eqno{(5.10)}
$$
We see that the $- \ln |\psi(z_k-z)|^2$ terms cancel with the
right-hand side, except for $k=j$, which we can trade with the
right-hand side for $\ln |\psi(z_i-z)|^2 = \ln|\psi(-\xi_i)|^2$ which
simplifies with the same term in the second line. We are thus left
with
$$
V \left\{ \sum_kN^{ik} \alpha_0^{(k)} + \sum_{k, k \neq i} \ln
|\psi(z_k-z_i)|^2 \alpha_0^{(k)} + \lim_{t \rightarrow 0}
\ln\left| {\psi(t) \over t}\right|^2 \alpha_0^{(i)} \right\} = V
\left\{ i \leftrightarrow j \right\}. \eqno{(5.11)}
$$
Thus we find our desired result that the purely momentum part of the vertex is
given by
$$
N^{ik} = \left\{ \matrix{\displaystyle{- \ln |\psi(z_k-z_i)|^2 \qquad
i \neq k} \cr
\displaystyle{- \lim_{t \rightarrow 0} \ln \left|
{\psi(t) \over t} \right|^2 \qquad i = k}} \right.
\eqno{(5.12)}
$$

\par
Now let us consider the $\alpha_m^k$ terms in (5.7). We make the
further assumption
$$
N_{nm}^{ik} = - {(-1)^n \over n!} {d^n \over dt^n}
\left. F_m(z_k-z_i+t) \right|_{t=0} \qquad k \neq i
$$
$$
N_{nm}^{ii} = - {(-1)^n \over n!} {d^n \over dt^n} \left. \left(
F_m(t) + {1 \over m} {1 \over (-t)^m} \right) \right|_{t=0},
\eqno{(5.13)}
$$
with analogous  equations for $\bar N_{nm}^{ik} $.
These relations also follow from the integrated
overlap relations. Therefore we can write
$$
V \left\{ - \sum_{m=1}^\infty \sum_{k, k \neq i} \alpha_m^{(k)}
\left( {(-1)^m \over m!} {d^m \over dt^m} \left. \left(
\ln\psi(z_i-z_k+t) \right) \right|_{t=0} + F_m(z_k-z) -
F_m(z_k-z_i) \right) \right.
$$
$$
\left. - \sum_{m=1}^\infty \alpha_m^{(i)} \left( F_m(-\xi_i) -
\lim_{t \rightarrow 0} \left( F_m(t) + {1 \over m} {1 \over
(-t)^m} \right) + {(-1)^m \over m!} \left. {d^m \over dt^m} \ln
\left({\psi(t) \over t}\right) \right|_{t=0} \right)\matrix{{}\cr
{} \cr {}}\right\}
$$
$$
= V \left\{ i \leftrightarrow j \right\},
\eqno{(5.14)}
$$
where, in the first line, we have used that
$$
\sum_{n=1}^\infty {(-1)^n \over n!} (\xi_i)^n {d^n \over dt^n}
\left. F_m(z_k-z_i+t) \right|_{t=0} = F_m(z_k-z) - F_m(z_k-z_i).
$$
And similarly for the second line. We see that the $F_m(z_k-z)$ terms
with $k \neq i,j$, simplify with the right-hand side, and the
remaining terms will cancel with the $F_m(-\xi_i)$ in the second
line. We are thus left with
$$
V \left\{ - \sum_{m=1}^\infty \sum_{k, k \neq i} \alpha_m^{(k)}
\left( {(-1)^m \over m!} {d^m \over dt^m} \left. \left(
\ln\psi(z_i-z_k+t) \right) \right|_{t=0} - F_m(z_k-z_i) \right)
\right.
$$
$$
\left. - \sum_{m=1}^\infty \alpha_m^{(i)} \left( - \lim_{t
\rightarrow 0} \left( F_m(t) + {1 \over m} {1 \over (-t)^m}
\right) + {(-1)^m \over m!} {d^m \over dt^m} \left. \ln
\left({\psi(t) \over t}\right) \right|_{t=0} \right) \matrix{{}\cr
{} \cr {}} \right\}
$$
$$
= V \left\{ i \leftrightarrow j \right\}.
\eqno{(5.15)}
$$
This equation is satisfied if
$$
F_m(z) = {1 \over m!} {d^m \over dt^m} \left. \ln \psi(-(z + t))
\right|_{t=0},
\eqno{(5.16)}
$$
and therefore
$$
N_{nm}^{ik} = \left\{ \matrix{\displaystyle{- {(-1)^m \over n! m!}
{d^{n+m} \over dt^{n+m}} \left. \ln \psi(z_i-z_k+t) \right|_{t=0}
\qquad i \neq k} \cr
\displaystyle{ -{(-1)^m \over n! m!} {d^{n+m} \over dt^{n+m}} \left.
\left( \ln \psi(t) - \ln t \right) \right|_{t=0} \qquad i = k}} \right.
\eqno{(5.17)}
$$
\par
For the open string we use the operator
$$
Q^\mu(z) = q^\mu - i \sqrt {2\alpha' } \alpha_0^\mu \ln |z| + i
\sqrt{2\alpha' } \sum_{n \neq 0} {1 \over n}  \alpha_n^\mu
z^{-n} .
\eqno{(5.18)}
$$
We now have $\alpha_0^\mu = \sqrt{2\alpha' }
p^\mu$, and using  the canonical commutation relation $[p^\mu, q^\nu] =
-i \eta^{\mu\nu}$, we can write
$$
q^\mu = -i \sqrt{2\alpha' } {\overleftarrow{\partial} \over
\partial \alpha_{0 \mu}},
\eqno{(5.19)}
$$
and we can therefore factor out the universal $i \sqrt{2\alpha' }$
factor and write
$$
Q^\mu(z) = i \sqrt{2\alpha' }
\left(-{\overleftarrow{\partial} \over \partial \alpha_{0 \mu}} -
\alpha_0^\mu \ln |z|^2 + \sum_{n \neq 0} {1 \over n} \alpha_n^\mu z^{-n}
 \right).
\eqno{(5.20)}$$ The operator $Q^\mu(z)$ has conformal weight one,
but it is not the same as the field $x^\mu$ for the open string
which depends on $z$ and $\bar z $. Rather, it is the field
$x^\mu(\tau,0)$ with $z=e^\tau$, extended by analytic continuation
into the upper half plane. And if the surface is doubled, also
into the lower half plane by reflection on the real axis. Of
course, before taking these steps we have mapped the field from
the string world-sheet to the upper half plane and rotated to
Euclidean space.
\par
The calculation of the purely momentum piece of the vertex proceeds much
as for the closed string if one just drops most of the anti-holomorphic
terms and terms containing $\bar \alpha_n$. We can write the vertex as
$$
V^{\rm open} = \prod_{k=1}^N {}_{(k)}\langle 0| \exp\left(-
\sum_{i,j} \left\{ {1\over2}\sum_{n,m=1}^\infty \alpha_n^{\mu {(i)}}
  N_{nm}^{ij}
\alpha_{m \mu}^{(j)} +  \sum_{n=1}^\infty \alpha_n^{\mu {(i)}}   N_n^{ij}
\alpha_{0 \mu}^{(j)}
\right\}\right)
$$
$$ \exp \left(-{1\over2} \sum_{i,j}\alpha_0^{\mu (i)}
  N^{ij} \alpha_{0 \mu}^{(j)} \right).
\eqno{(5.21)}
$$
We are using the same symbols for the coefficients in the open string
vertex, the reader will know from the context which string is being
considered. We begin from the overlap relations
$$
V Q^{\mu (i)}(\xi_i) = V Q^{\mu (j)}(\xi_j), \eqno{(5.22)}
$$
and  assume that $  N_n^{ik}$ has the form
$$
  N_n^{ik} = - {(-1)^n \over n!} \left. {d^n \over dt^n}
\left( \ln \psi(z_k-z_i+t) \right) \right|_{t=0} \qquad k \neq i
$$
$$
  N_n^{ii} = - {(-1)^n \over n!} \left. {d^n \over dt^n} \left( \ln
\psi(t) - \ln t \right) \right|_{t=0},
\eqno{(5.23)}
$$
Comparing the $\alpha_0$ terms in equation (5.22)  we find
that  the purely momentum part of the vertex is
given by
$$
  N^{ik} = \left\{ \matrix{\displaystyle{- \ln |\psi(z_k-z_i)| \qquad
i \neq k} \cr
\displaystyle{- \lim_{t \rightarrow 0} \ln \left|
{\psi(t) \over t} \right| \qquad i = k}} \right.
\eqno{(5.24)}
$$
While comparing the $\alpha_n, n>0$  and assuming that
$$
  N_{nm}^{ik} = - {(-1)^n \over n!} {d^n \over dt^n}
\left.   F_m(z_k-z_i+t) \right|_{t=0} \qquad k \neq i
$$
$$
  N_{nm}^{ii} = - {(-1)^n \over n!} {d^n \over dt^n} \left. \left(
  F_m(t) + {1 \over m} {1 \over (-t)^m} \right) \right|_{t=0},
\eqno{(5.25)}
$$
we find that  equation (5.22) is satisfied if
$$
  F_m(z) = {1 \over m!} {d^m \over dt^m} \left. \ln \psi(-(z + t))
\right|_{t=0},
\eqno{(5.26)}
$$
In fact equations (5.23) and (5.25) follow from the integrated overlap
equations. As a result, we find that
$$
  N_{nm}^{ik} = \left\{ \matrix{\displaystyle{- {(-1)^m \over n! m!}
{d^{n+m} \over dt^{n+m}} \left. \ln \psi(z_i-z_k+t) \right|_{t=0}
\qquad i \neq k} \cr
\displaystyle{ -{(-1)^m \over n! m!} {d^{n+m} \over dt^{n+m}} \left.
\left( \ln \psi(t) - \ln t \right) \right|_{t=0} \qquad i = k.}} \right.
\eqno{(5.27)}
$$


\bigskip
{\bf {6. One Loop Open and Closed Arbitrary String Scattering}}
\medskip
In references [13,14] the planar open string multi-string
scattering vertex and measure were computed. In the first
subsection we first calculate the one loop multi-string vertex for
closed string scattering and in the next subsection the measure.
Finally we do the same for non-planar open string scattering.
\medskip
{\bf {6.1 The nonzero mode part of the Closed Vertex}}
\medskip
We use the Fuchsian representation of the Riemann surface
as explained at the beginning of section four, and we will adopt the local
coordinates $\xi_i=z-z_i$. For the closed string we take $w^\prime $
arbitrary but
$2w=1$, however, as similar calculations are useful for the open string
we will sometimes keep factors of $w$ explicitly rather than set it
equal to ${1\over 2}$. As explained in section two we may use the
integrated overlap identities of equation (2.15) for the conformal
operator $\partial x^\mu$  to determine much of the string vertex.  These
identities require scalar functions
$\varphi$
 which are  defined on the torus and so in our representation of the
torus obey
$$
\varphi(z)=\varphi(z+2w),\ \ \varphi(z)=\varphi(z+2w^\prime),
\eqno(6.1.1)
$$
and analytic except for possible poles at the Koba-Nielsen points.  The
identities are most useful for functions that have poles at only one
Koba-Nielsen point. A complete set of such functions are given by the
Weierstrass function ${\cal P}$ and its derivatives, and so we take
$$\varphi_n(\xi_i)={(-1)^n\over (n)!}{\cal P}^{(n-2)}(\xi_i)
=\cases {{1\over n(\xi_i)^n}+N_{n}^{ii}+\sum
_{m=1}N_{nm}^{ii}(\xi_i)^m\cr
N_{n}^{ij}+\sum_{m=1}N_{nm}^{ij}(\xi_j)^m}\ \ n\ge 2 .
\eqno(6.1.2)$$
The  overlap equation is then given by
$$
V\left({1\over n} \alpha_{-n}^{ \mu (i)}+\sum_j
N_{n}^{ij}\alpha_0^{\mu (j)}+\sum_{j}\sum_{m=1}^\infty
N_{nm}^{ij}\alpha_m^{\mu (j)}
  \right)=0,\ \ n\ge 2.
\eqno(6.1.3)
$$
We note that, for $n \geq 2$
$$
N_{nm}^{ij} = {(-1)^n \over n!m!} \left.{d^{(n+m-2)} {\cal P}(z_j-z_i+t) \over
dt^{(n+m-2)}}\right|_{t=0}=
{(-1)^m \over n!m!} \left. {d^{(n+m-2)} {\cal P}(z_i-z_j+t) \over
dt^{(n+m-2)}}\right|_{t=0},\ i\neq j
$$
$$
N_{nm}^{ii} = \left. {(-1)^n \over n!m!} {d^{(n+m-2)} ({\cal P}(t)-{1\over
t^2})
\over dt^{(n+m-2)}}\right|_{t=0}
$$
$$
N_n^{ij} = {(-1)^n \over n!} \left. { d^{(n-2)}{\cal P}(z_j-z_i+t)\over
dt^{(n-2)}}\right|_{t=0},\  i\neq j
$$
$$
N_n^{ii} = {(-1)^n \over n!} \left. {d^{(n-2)} ({\cal P}(t) - t^{-2})
\over dt^{(n-2)}} \right|_{t=0}.
\eqno(6.1.4)
$$
\par
There is  no scalar function with a
pole of order one at a single point, and so at first sight we do not have
an integrated overlap identity that can determine the dependence of the
vertex on $\alpha_{1}^\mu$. The solution to
 this difficulty is to consider
functions which are not periodic, but shift by constants under the
change of periods, namely a  function such that
$$
\varphi(z+2w)=\varphi(z)+c_1,\ \
\varphi(z+2w^\prime)=\varphi(z)+c_2
\eqno(6.1.5)$$
We now consider the contour integral of $V\partial x^\mu\varphi$ around
the boundary of the parallelogram that describes the torus. The
contributions from opposing sides no longer cancel as $\varphi$ is
no longer defined on the torus and one finds, instead of equation
(2.15),  that
$$
\sum_j \oint d\xi_j V\varphi\partial
x^{\mu(j)}(\xi_j)=c_2\int_{B_1}V\partial x^{\mu}dz
+c_1\int_{A_2}V\partial x^{\mu}dz ,
\eqno(6.1.6)
$$
where the paths $B_1$ and $A_2$ are shown in figure 1. On the
right hand side $V\partial x^{\mu}dz$ stands for $V\partial
x^{\mu(k)}(\xi_k)d\xi_k$, which is independent of which string
Hilbert space one uses by virtue of equation (2.5) using the
operator $\partial x^{\mu}$. It should not matter which function
$\varphi$ one takes as long as it obeys equation (6.1.5).
Functions of this type, with a  pole of order one at the same
place, differ by a constant times $z$.
 Taking
$\varphi=z$ in equation (6.1.6) we find that provided
$$
 w^\prime\int_{B_1}V\partial x^{\mu} dz
+w\int_{A_2}V\partial x^{\mu}dz={1\over 2}V\left(\sum_j z_j
\alpha_0^{\mu (j)}+\sum_j \alpha_1^{\mu (j)}\right),
\eqno(6.1.7)$$
we will indeed find the same result no matter what
function $\varphi$ we take.
\par
Equation (6.1.7)  does not determine the  integrals along the
$B_1$ and $A_2$ segments of the parallelogram, but as we will
argue later, the unique solution taking into account modular
invariance, is given by
$$
\int_{B_1}V\partial x^{\mu} dz={\bar w\over (w^\prime-\bar
w^\prime)}V\sum_j \alpha_1^{(j)}-V\sum_j{(w\bar z_j-\bar w z_j)\over
(w^\prime-\bar w^\prime)}
\alpha_0^{\mu (j)}
$$
$$
\int_{A_2}V\partial x^{\mu}dz =-{\bar w^\prime\over (w^\prime-\bar
w^\prime)}V\sum_j \alpha_1^{(j)}+V\sum_j{(w^\prime\bar z_j-\bar
w^\prime z_j)\over (w^\prime-\bar w^\prime)} \alpha_0^{\mu (j)}.
\eqno(6.1.8)$$
We observe that under the exchange
$w\leftrightarrow w^\prime$ the two integrals are exchanged with
minus sign provided we recognise that $w^\prime-\bar
w^\prime=2i{\rm Im} \ w^\prime= i{\rm Im} \tau$ is $i$ times the
area of the fundamental region and is unchanged in this process.
\par The function
$$
\varphi_1(z)= \zeta(z)-{\eta\over w}z ,
\eqno(6.1.9)$$
 obeys equation (6.1.5) with $c_1=0$ and $c_2=-{i\pi\over w}$. In this
equation $\zeta(z)$ is the zeta function which is defined to obey
$-{d\zeta(z)\over dz}={\cal P}(z)$ and has a first order pole with residue
one.  It is not a periodic function, but obeys
$$
\zeta(z+2w)=\zeta(z)+2\eta,\ \ \ \zeta(z+2w^\prime)=\zeta(z)+2\eta^\prime
\eqno(6.1.10)$$
where $\eta$ and $\eta^\prime$ are independent of $z$. See appendix A for
more details.  Substituting $\varphi_1$  into
equation (6.1.6)   and using equation (6.1.8) we find that
$$
V\left(\alpha_{-1}^{ \mu (i)}+\sum_{j} \sum_{m=1}^\infty\tilde N_{1
m}^{ij}\alpha_m^{\mu (j)} + \sum_{j} \tilde N_{1}^{ij}\alpha_0^{\mu
(j)} \right)
$$
$$=-{i\pi\over w}V \left( {\bar w\over (w^\prime-\bar
w^\prime)}\sum_j \alpha_1^{(j)}-\sum_j{(w\bar z_j-\bar w z_j)\over
(w^\prime-\bar w^\prime)}
\alpha_0^{\mu (j)} \right).
\eqno(6.1.11)$$
\par
The coefficients in this equation are just those found by expanding
$\varphi_1$ around the Koba-Nielsen points in terms of the local
coordinates $\xi_i$, indeed
$$
\varphi_1(z-z_i)= \cases {{1\over \xi_i}+\sum
_{m=1}\tilde N_{1m}^{ii}(\xi_i)^m\cr
\tilde N_{1}^{ij}+\sum_{m=1}\tilde N_{1m}^{ij}(\xi_j)^m}.
\eqno(6.1.12)$$
Clearly, the coefficients are given by
$$
\tilde N_{1m}^{ij}= {1\over m!} \left. {d^m
\varphi_1(z_j-z_i+t)\over dt^m}\right|_{t=0} \ \ , \quad \tilde
N_{1}^{ij}=\varphi_1(z_j-z_i), \ \ i\not= j
$$
$$
\tilde N_{1m}^{ii}={1\over m!}\left.{d^m \over
dt^m}\left(\varphi_1(t)-{1\over t}\right)\right|_{t=0}
\eqno(6.1.13)
$$
\par
From the integrated overlap equations (6.1.3) and (6.1.11) we read
off the vertex to be
$$
V^{\rm closed} =  \prod_{k=1}^N {}_{(k)}\langle 0| \exp \left(-
\sum_{i,j} \left\{ {1\over2}\sum_{n,m=1}^\infty \alpha_n^{\mu {(i)}}
N_{nm}^{ij}
\alpha_{m \mu}^{(j)} +  \sum_{n=1}^\infty \alpha_n^{\mu {(i)}} N_n^{ij}
\alpha_{0 \mu}^{(j)} \right. \right.
$$
$$
\left. \left. + \left(N_{nm}^{ij}\to \bar N_{nm}^{ij},\
N_n^{ij}\to \bar N_n^{ij},\ \alpha_n^{\mu (k)}\to \bar
\alpha_n^{\mu (k)}\right) \matrix{{}\cr{}}\right\}
\matrix{{}\cr{}\cr{}}\right) \exp \left(-{1\over2}
\sum_{i,j}\alpha_0^{\mu (i)} N^{ij} \alpha_{0 \mu}^{(j)} \right).
\eqno{(6.1.14)}
$$
The coefficients $N_{nm}^{ij}$ and $N_{n}^{ij}$ for $n\ge 2$ are given in
equation (6.1.4),  while the remainders are
$$
 N_{1m}^{ij}=\tilde  N_{1m}^{ij} +{\pi i\over
(w^\prime-\bar w^\prime)}\delta_{1,m},\ \
 N_{1}^{ij}=\tilde  N_{1}^{ij}-{2\pi  {\rm Im} (z_j-z_i)\over
(w^\prime-\bar w^\prime)}
\eqno(6.1.15)$$
The additions to the latter coefficients, and so to the vertex, arise from the
contributions on the right hand side of equation (6.1.11) which in turn
are a consequence of the non-periodic nature of $\varphi_1$.
\par
The zero mode piece can be read off from section five equation
(5.12) in terms of the coefficients that appear in the vertex. In
deriving this result,  we made some assumptions about the
coefficients and we now show that these are a consequence of the
results above. Let us consider the function
$$
\ln \psi(z) = \left( \ln \sigma(z) -{\eta\over 2w}z^2 \right) +
c.c -{2\pi\over {\rm Im} \tau}({\rm Im} z)^2 =\ln {\theta_1
(v)\over \theta_1^\prime(0)}+c.c -{2\pi\over {\rm Im} \tau}({\rm
Im} z)^2. \eqno(6.1.16)
$$
The $\sigma$ function is
defined by
$$
\zeta(z) = {d \over dz} \ln \sigma(z) \qquad , \qquad {\cal P}(z) =
- {d \zeta(z) \over dz}
\eqno(6.1.17)
$$
and in the last step of equation (6.1.16) we have used the
relation between the $\sigma$ function and the $\theta_1$ function
to give an alternative expression. The argument of the theta
functions is $v={z\over 2w}$, but it also depends on $\tau$. We
observe that the function is not a holomorphic function of $z$
only  as a consequence of the last term. As such, when taking the
derivatives we use the expression
$$
{d\over d z}={1\over 2}\left({d\over d z_1}-i{d\over d z_2}\right)
\eqno(6.1.18)
$$
where $z=z_1+iz_2$. Calculating the coefficients using equations
(5.8) and equations (5.17) we do find precisely the results of
equations (6.1.4), (6.1.13) and (6.1.15). This is rather obvious
for terms which have  more than two derivatives as the last two
terms in equation (6.1.16) do not contribute. This only leaves one
and two derivatives which is easily carried out. One finds that
the non-holomorphic dependence of $\ln \psi$ arises from the
 the contour integral along  $B_1$
of equation (6.1.11). Thus we find, as for all string vertices, that it is
determined by only one function $\ln \psi$. One can think of this as
 a consequence of the definition of the vertex in terms of overlap
relations.
\medskip
{\bf {6.2 The measure for the Closed Vertex}}
\medskip
As explained in section two the measure is determined by the
decoupling of zero norm null physical states. In the simple
coordinate system in which we are working, i.e. $ \xi_i=z-z_i$, at
genus one, and indeed for all genus, one finds that
$$
VL_{-1}^{(i)}={\partial V\over \partial z_i}
\eqno(6.2.1)
$$
The explanation is the same as for the case of tree level in
equation (2.49). Following the discussion below this equation,
that is equations (2.50) and (2.51) we conclude that the measure
$f_m$ is  independent of $z_i$.
\par
Decoupling of the zero norm physical states at higher levels requires us
to understand how a conformal
transformation changes the moduli i.e. $w^\prime \to w^\prime +\epsilon$
as we have taken $2w=1$. As such,  we need an $f$ in equation (2.25) such
that
$$f(z)=f(z+2w),\ \ \ f(z)=2+f(z+2w^\prime)
\eqno(6.2.2)$$
Such a function is given by
$$
f(z)={2w\over i\pi}\left(
 \zeta(z)-{\eta\over w}z\right)
\eqno(6.2.3)$$
Substituting this function with argument $z-z_i$ into equation (2.27) we
find that
$$
{i\pi\over 2 w}{\partial V\over \partial
w^\prime}=V\left(L_{-2}^{(i)}-{\eta\over w}L_{0}^{(i)}+c \right)
+\sum_{j,j\not=i} \left(\zeta(z_j-z_i)-{\eta\over
w}(z_j-z_i)\right)VL_{-1}^{(j)}
$$
$$
+\sum_{j,j\not=i} \left( {d \zeta(z_j-z_i)\over dz_j}-{\eta\over
w}\right) VL_{0}^{(j)}+\ldots ,
\eqno(6.2.4)
$$
where $+\ldots$ means terms with $L_n$ for $n\ge 1$. Taking the
scalar product of this equation with the vacuum state, with zero
momentum,  on all string Hilbert spaces and normalizing the vertex
as in equation (5.1) we find that
$$
c={D\over 2} N^{11}_{11}={D\over 2}\left(-{\eta\over w}+{\pi i\over 2 w
(w^\prime-\bar w^\prime)}\right)
\eqno(6.2.5)$$
where $D$ is the dimension of space-time and we have used equation
(6.1.15)
\par
Applying the zero norm state
$(L_{-2}^{(i)} +{3\over 2}(L_{-1}^{(i)})^2)|\Omega^\prime \rangle_i$ on
the
$i$th string Hilbert space and
 arbitrary physical states $|\chi_k\rangle$ to all the other Hilbert spaces
to the integrated vertex
$$W = \int \prod_{j=1}^{N-1} d^2z_j d^2w^\prime f_m V
\eqno(6.2.6)
$$
we find that
$$
W|\chi\rangle_{(1)} \ldots |\chi\rangle_{(N)} =
$$
$$
= \int \prod_{k=1}^{N-1} d^2z_k
d^2w^\prime f_m \left( {i\pi\over 2 w}{\partial V\over \partial
w^\prime} +V\left({\eta\over w}L_{0}^{(i)}-c\right)
-\sum_{j,j\neq i} \left(\zeta(z_j-z_i)-{\eta\over
w}(z_j-z_i)\right)VL_{-1}^{(j)}
\right.
$$
$$ \left. -\sum_{j, j\neq i} \left({d \zeta(z_j-z_i)\over dz_j}-{\eta\over
w}\right)VL_{0}^{(j)} \right) |\chi\rangle_{(1)} \ldots
|\Omega\rangle_{(i)} \ldots |\chi\rangle_{(N)} . \eqno(6.2.7)
$$
Using equation (6.2.1) and integrating by parts we find that all the
terms containing the $\zeta$ and terms in the same brackets
vanish. Integrating by parts we then conclude that
$$
{\partial \ln  f_m\over \partial \tau}={(D-2)}{\eta\over \pi
i}  -{D\over 2}{1\over 2(w^\prime-\bar w^\prime)}
\eqno(6.2.8)
$$
\par
To solve this equation we must recall some facts about theta functions;
$\theta_j, j=1,\ldots ,4$ are  functions of the variable $v$ and also
depend on $\tau$, they are sometimes   written as
$\theta_j(v|\tau)$. These functions  for an arbitrary parameterisation
of the torus, i.e. arbitrary $w$ and $w^\prime$, obey the equation
$$
{\partial^2 \theta_j(v|\tau)\over \partial v^2}= 4\pi i
{\partial \theta_j(v|\tau)\over \partial \tau}
\eqno(6.2.9)$$
Now  $\eta$ can be expressed as [24]
$$
\eta =-{1\over 12 w} {\left. {\partial^3 \theta_1(v|\tau)\over
\partial v^3}\right|_{v=o} \over \left. {\partial
\theta_1(v|\tau)\over \partial v}\right|_{v=o}} .\eqno(6.2.10)
$$
Using equation (6.2.9) we find that
$$
\eta =-{\pi i\over 3 w} {\partial \over \partial \tau} \left( \ln
\left. {\partial \theta_1(v|\tau)\over \partial v} \right|_{v=o} \right)
\eqno(6.2.11)$$
Examining the explicit expression for  $\theta_1 (v|\tau)$ given in
the appendix A we conclude that
$$
\left. {\partial \theta_1(v|\tau)\over \partial v} \right|_{v=o}
= 2 \pi q^{{1\over 4}} \left( \prod_{n=1}^\infty (1-q^{2n}) \right)^3
\eqno(6.2.12)$$
where  $q=\exp (\pi i \tau)$.
\par
Substituting equation (6.2.11) into equation (6.2.8) and using equation
(6.2.12) we find that the measure is given by
$$
 f_m= {1\over | \left( \Pi_{n=1}^\infty (1-q^{2n}) \right)^{D-2} |^2}{1\over
|q|^{{D-2\over 6}}}{1\over ({\rm Im} \tau)^{{D\over 2}}}
\eqno(6.2.13)$$
which is indeed the know result. The final  contribution containing
${\rm Im} \tau$ comes from the last term in equation (6.2.8) arising from the
constant in equation (6.2.5), and this in turn can be traced to the right
hand side of equation (6.1.11), which in turn arises from the $B_1$ contour
integral in equation (6.1.8). We note that this contribution  depends on
$\tau$ and $\bar \tau$ so one must differentiate with respect to
$\tau$ using an expression analogous to that of equation (6.1.18). Hence
the non-holomorphic dependence in the measure arises from the use of
non-periodic functions in the integrated overlaps.
\par
The modular invariance of the measure requires just such a contribution
and one could have taken the  $B_1$ contour integral to be given
by arbitrary expressions times $\alpha_1$ and $\alpha_0$ oscillators. As is
apparent from equation (6.2.5), the former is the one which contributes
to the measure through the constant
$c$ which first appears in  equation  (6.2.4). One would then find that
the $\alpha_1$  term in the
 $B_1$ contour integral takes the value it does in equation (6.1.8).
The part of the $B_1$ contour integral which contains $\alpha_0$
contributes to the non-holomorphic part contained in the purely zero mode
contribution of the vertex, and again it is fixed by direct
comparison with the known result, or demanding modular invariance.

\medskip
{\bf {6.3 Non-planar one loop open string scattering}}
\medskip
The calculation of one loop open string scattering has many steps
in common with that for closed strings given in  the previous two
sub-sections. The main difference arises from the fact that the
open strings sweep out a surface with boundaries and it is on
these boundaries that the open strings are emitted, i.e. the
Koba-Nielsen points are on the boundaries.   At one loop the
world-sheet is a cylinder which has two boundaries from which the
open strings may be emitted. One can think of the cylinder as
being due to an intermediate closed string. The calculation is
rather sensitive to how the strings are distributed between the
two boundaries.  If all  the emitted strings  are  on the same
boundary, the world-sheet can be drawn on a plane (the external
states being drawn as semi-infinite strips), hence this diagram is
called {\it planar}. This case has been already computed in
[13,14]. We want here to calculate the non-planar contribution
which has strings on both boundaries (and thus cannot be drawn on
a plane). These non-planar contributions are especially relevant
to the main subject of this paper as   it was shown [1,4] that, at
high energy, it is these diagrams that dominate the four tachyon
scattering amplitude.
\par
We double the open string world-sheet to construct a torus which,
using  the Fuchsian representation, is represented as a rectangle
in the complex plane, with corners at $z= -w'$, $-w'+2w$, $w'+2w$
and $w'$, where $w$ is real and $w'$ is purely imaginary.  The
boundaries of the open string $I$ and $I'$ are the segments inside
the rectangle with ${\rm Im} z = 0$ and ${\rm Im} z= {\rm Im} w'$
respectively. We adopt the choices
$$
2 w =- \ln {u} \qquad , \qquad w' = i \pi.
\eqno(6.3.1)
$$
As $w^\prime$ and $w$ are purely imaginary and real respectively, the
fundamental region is a rectangle for the open string.  As
$u\le 1$,
$-\ln u$ is a positive number. In the old literature the quantity $u$
is often called $w$, but following Bateman we are already using $2w$ for
one of the periods of the torus.

\par
We first construct the vertex whose general form is given in equation
(5.21). The
integrated overlaps  with periodic functions, which obey
equation (6.1.1), lead to equation (6.1.2) and then equations (6.1.3)
and (6.1.4). Thus the coefficients $N_{nm}^{ij}, \ N_{n}^{ij}, \ n\ge 2$
 in the open string vertex are given by equation (6.1.4).
\par
As for the
closed case there are no  periodic functions with a first order pole at a
given point on a torus, and we use functions which are shifted by constants
when their arguments change by the periods of the torus, i.e. equation
(6.1.5).  The integrated overlap equation for such functions is given in
equation (6.1.6) and
 consistency then requires us to
adopt equation (6.1.7) that is
$$
 w^\prime\int_{B_1}dzV\partial x^{\mu}
+w\int_{A_2}dzV\partial x^{\mu}={1\over 2}V\left(\sum_j z_j
\alpha_0^{\mu (j)}+\sum_j \alpha_1^{\mu (j)}\right) .
\eqno(6.3.2)
$$
This equation does not determine the individual contour integrals, but
the general solutions can be written in the form
$$
\int_{B_1}dz V \partial x = b^{(2)} \sum_j \alpha_1^{(j)} +
\sum_j b_j^{(1)} \alpha_0^{(j)}.
\eqno(6.3.3)
$$
We will adopt the value
$$
\int_{B_1}dz V \partial x = {1 \over 2} \sum_{j \in I'} \alpha_0^{(j)} =
{1 \over 2 \pi} \sum_j {\rm Im} (z_j - z_i) \alpha_0^{(j)}.
\eqno(6.3.4)
$$
This is a natural result; the one loop  world-sheet can be seen
either as an open string going around the cylinder, or as a closed
string propagating from one boundary to the other. In this second
picture, the integral on the left-hand side of (6.3.4) measures
the momentum of the closed string, which in our case is equal to
the total momentum going in the $I'$ boundary. The factor of ${1
\over 2}$ comes from the relation between $p^\mu$ and
$\alpha_0^\mu$, which is different for closed and open strings, as
can be seen from (5.3) and (5.18).  The value of the $B_1$ contour
integral in equation (6.3.3) feeds into various non-holomorphic
corrections to the purely zero mode part of the vertex and the
measure as we will explain. While we can not, as for the closed
string use modular invariance, we can compare the results with
that found long ago for tachyon scattering by sewing tree level
three vertices. From this   we can deduce the values of $b^{(2)}$
and $ b_j^{(1)}$. This is in fact the route we have taken, but for
simplicity we will use the correct values as given in equation
(6.3.4)
\par
Proceeding with the calculation of the vertex,  we
substitute
$\varphi_1(z) =
\zeta(z) - {\eta
\over w} z$ into equation (6.1.6) and use equation (6.3.4) to find
$$
V\left(\alpha_{-1}^{ \mu (i)}+\sum_{j} \sum_{m=1}^\infty\tilde N_{1
m}^{ij}\alpha_m^{\mu (j)} +\sum_{j} \tilde N_{1}^{ij}\alpha_0^{\mu
(j)}\right) = -{i\over2 w}V\sum_j {\rm Im} (z_j - z_i) \alpha_0^{(j)},
\eqno(6.3.5)
$$
where the $\tilde{N}$'s are given  in equation (6.1.13).  The
coefficients in the  vertex as calculated from equation (6.3.5) are given
by
$$
N_1^{ij} = \tilde{N}_1^{ij} + {i \over 2 w} {\rm Im} (z_j-z_i),\ \
\tilde{N}_{1m}^{ij}={N}_{1m}^{ij} .
\eqno(6.3.6)
$$
We note that we only get a contribution from the non-periodicity of
$\varphi_1$ when $z_j$ and $z_i$ are on different boundaries.
\par
We assumed in the derivation of the purely zero mode piece that the
coefficients in the vertex are the derivatives of a single function. We
now show that this is indeed the case for the above vertex. In doing so
we will treat the cases of whether
$z_j$ and $z_i$ are on the same boundary or not differently.  If they are
on the same boundary, the non-periodic
nature of $\varphi_1$ has no effect and so
${\rm Im} (z_j-z_i) = 0$ and $N_1^{ij} = \tilde{N}_1^{ij}$.  As the reader
will readily verify, for these
$i$ and $j$ all the coefficients are given by equations (5.23) and
(5.26)  if we take
$$
\ln \psi(z) = \ln \sigma(z) - {\eta \over 2 w} z^2 -2\ln w=
\ln {\theta_1(v) \over \theta'_1(0)},
\eqno(6.3.7)
$$
where $v= {z \over 2 w}$.
The zero mode coefficient is then given by equation (5.24). These are the
same as for the  case of the planar string.
\par
If $z_i$ and $z_j$ are on different  boundaries, we have ${\rm Im}
(z_j-z_i) = \pm \pi$ and so from equation (6.3.6) we have
$N_1^{ij} = \tilde{N}_1^{ij} \pm {i\pi\over 2 w}$. However, in
this case, two of the Koba-Nielsen points are on the line ${\rm
Im} z=0$ while the other two are on the line ${\rm Im} z=\pi$
within the fundamental region. As it is desirable to integrate
over the real variable ${\rm Re} z_j$ we will consider functions
of ${\rm Re} z_j$ and absorb the imaginary contributions to the
Koba-Nielsen points in the definition of the function. Taking
these two requirements into account, the  function that leads to
the coefficients of the vertex using  equations (5.23) and (5.26)
is $\psi_T(z)$, which is given by
$$
\ln \psi_T = \ln {\theta_4( {\rm Re} v) \over \theta_1'(0)} .
\eqno(6.3.8)
$$
To see that this is the case we need the identity [24]
$$
\ln  {\theta_4(z)\over \theta_4(0)}=\ln \sigma(z+w^\prime) -\eta^\prime z
-{\eta\over 2 w}z^2 - \ln \sigma(w')
\eqno(6.3.9)$$
we recall that  $v = {z \over 2 w}$.
Differentiating equation (6.3.8) and using the above
identity we find that
$$
{d \over dt} \ln \psi_T({\rm Re} z_j-{\rm Re} z_i+t)|_{t=0} =
\zeta({\rm Re} z_j - {\rm Re} z_i + i\pi) -\eta^\prime
- {\eta \over w} ({\rm Re} z_j - {\rm Re} z_i )
$$
$$
=\zeta({\rm Re} z_j - {\rm Re} z_i \pm i\pi) -
{\eta \over w} ({\rm Re} z_j - {\rm Re} z_i \pm i \pi) \pm {i \pi \over 2 w}
\eqno(6.3.10)$$
where in deriving the last equation we have used the Legendre relation
and $\zeta(z+2w^\prime)=\zeta(z)+2\eta^\prime$ to derive the
relationship with the lower sign.
Hence, we find that
$$
{d \over dt} \ln \theta_4({\rm Re}\ v_j-{\rm Re}\ v_i+t) |_{t=0}=
\varphi_1({\rm Re}\ z_j-{\rm Re}\ z_i\pm i\pi) \pm  {i \pi \over 2
w}. \eqno(6.3.11)
$$
In the argument of $\varphi_1$ we recognise the positions of the
Koba-Nielsen points as they are on the rectangle, i.e we find in
this equation $\varphi_1(z_j- z_i)$ and the last term is just that
required to get the correct coefficient $N_1^{ij} $ in equation
(6.3.6). This last term disappears when taking further derivatives
and  one recovers all the other coefficients in the vertex  given
above using equations (5.23)  and (5.26).
\par
In summary, we have
$$
N^{ij} = \left\{ \matrix{\displaystyle{\ln \psi(z_j-z_i) \quad i, j
\ {\rm on \ the \ same \ boundary}} \cr
\displaystyle{\ln \psi_T({\rm Re}\ z_j-{\rm Re}\ z_i) \quad i, j
\ {\rm on \ different \ boundaries}}} \right.
\eqno(6.3.12)
$$
with
$$
\ln \psi(z) = \ln {\theta_1(v) \over \theta_1'(0)}
$$
$$
\ln \psi_T(z) = \ln {\theta_4(v) \over \theta_1'(0)}
\eqno(6.3.13)
$$
where $\theta_j'(0)={\partial\theta_j (v|{\tau})\over
\partial v} |_{v=0}$. We note that if $i,j$ are on the same boundary $z_j-z_i$
is
real.
\par
The calculation of the measure for the open string has many steps
in common with that for the closed string given in section 6.2.
From the decoupling of $L_{-1}^{(j)}|\Omega\rangle$, we  conclude
that the measure $f_m$ is independent of $z_j$. For the decoupling
of the zero norm physical states at higher levels, we must
understand how to change to dependence of the vertex on its
modulus using a conformal transformation while $w' = i \pi$ is
fixed and we change $u\to u(1-\epsilon)$. Under this change $w
\rightarrow w + {\epsilon \over 2}$ which is implemented by the
conformal transformation with  an $f$ in equation (2.25) such that
$$
f(z)=f(z + 2w'),\ \ \ f(z)=1+f(z + 2w).
\eqno(6.3.14)
$$
Such a function is given by
$$
f(z) = - \left(\zeta(z)-{\eta' \over i \pi}z\right) \eqno(6.3.15)
$$
Substituting this function with argument $z-z_i$ into equation (2.27) we
find that
$$
u{ \partial V\over \partial
u} = V \left(L_{-2}^{(i)}-{\eta' \over w'}L_{0}^{(i)}+c \right)
+\sum_{j, j \neq i} \left(\zeta(z_j-z_i)-{\eta' \over
w'}(z_j-z_i)\right)VL_{-1}^{(j)}
$$
$$
+\sum_{j, j \neq i} \left({d \zeta(z_j-z_i)\over dz_j}-{\eta' \over
w'}\right)VL_{0}^{(j)}+\ldots
\eqno(6.3.16)
$$
Taking the scalar
product of this equation with the vacuum state, with zero momentum,  on
all string Hilbert spaces
and normalizing the vertex as in equation (5.21) we find that
$$
c={D\over 2} N^{11}_{11}= - {D\over 2} {\eta\over w}.
\eqno(6.3.17)
$$
\par
Applying the zero norm state
$(L_{-2}^{(i)} +{3\over 2}(L_{-1}^{(i)})^2)|\Omega^\prime \rangle_i$ on
the
$i$th string Hilbert space and
 arbitrary physical states $|\chi_k\rangle$ to all the other Hilbert spaces
to the integrated vertex, we find that
$$
\int \prod_{k=1}^{N-1} dz_k du \left(
f_m \left(u{\partial V\over \partial u}
+V{\eta'\over w'}L_{0}^{(i)}-Vc\right)
-\sum_{j,j\neq i} \left(\zeta(z_j-z_i)-{\eta'\over w'}(z_j-z_i)
\right) VL_{-1}^{(j)} \right.
$$
$$
\left. -\sum_{j,j\neq i} \left({d \zeta(z_j-z_i)\over
dz_j}-{\eta'\over w'}\right)VL_{0}^{(j)}\right) |\chi\rangle_{(1)}
\ldots |\Omega\rangle_{(i)} \ldots |\chi\rangle_{(N)} = 0.
\eqno(6.3.18)
$$
As for the closed string,  the terms containing the
$\zeta$ and other terms in the same brackets vanish by integration by
parts. Integrating by parts with respect to $u$, and using Legendre's
relation we then conclude that
$$
u{\partial \ln  f_m\over \partial u} = {(D-2)\over 2} {\eta\over w}
+{1\over 2w}-1.
\eqno(6.3.19)
$$
\par
To solve this equation we require an expression for $\eta$ in terms of
$u$.  Using the well known relation
$$
\left. {\partial\theta_1 (v|{a\tau+b\over c\tau+d })\over \partial v}
\right|_{v=0}
=e(c\tau+d)^{{3\over 2}} \left. {\partial\theta_1 (v|{\tau})\over
\partial v} \right|_{v=0}
\eqno(6.3.20)$$
where $e$ is a constant whose value will not be important in what
follows, and equation (6.2.12) we find that
$$
q^{{1\over 12}} \prod_{n=1}^\infty (1-q^{2n})=u^{{1\over 24}}
\left({ -\ln u \over 2\pi}\right)^{{1\over 2}}\prod_{n=1}^\infty(1-u^{n}).
\eqno(6.3.21)$$
Using equation (6.2.11) we then find that the measure for one loop
non-planar scattering is given by
$$
f_m = {1 \over {u} {u}^{D-2 \over 24}
 (\prod_{n=1}^\infty(1-u^{n}))^{D-2}
 (-\ln u)^{D/2}} .
\eqno(6.3.22)
$$
This displays the expected reduction in  degrees of freedom from $D$ to
$D-2$.

\bigskip
{\bf {7. Scattering at Higher Genus and Discussion}}
\medskip
It was argued in references [2,3,4] that the four tachyon closed
string scattering is dominated, at high energy and fixed angle, by a
world-sheet surface of the form
$$
y^N= {(\varsigma-\varsigma_1)(\varsigma-\varsigma_2)\over
(\varsigma-\varsigma_3)(\varsigma-\varsigma_4)} \equiv (Q(\varsigma))^N
\eqno(7.1)$$
This is a Riemann sphere with two branch cuts of degree $N$, and so is in
effect a  set of $N$ Riemann spheres connected by the two branch cuts
which are at the same points for each sheet.  It is a surface of genus
$g=N-1$. The strings are emitted from the points at the beginning and end
of the branch cuts, i.e. the Koba-Nielsen points are $\varsigma_i,\
i=1,2,3,4$.  The authors of references [2,3,4] argued that the tachyon
scattering contains an exponential factor, bilinear in
the momenta, which is
${1\over N}$ times the tree level result. This essentially results from
the fact that the Green's function is identical on all the sheets and
there are $N$ of them.
\par
We will now use the group theoretic method to derive some of the string
vertex for the four closed string scattering at arbitrary genus, and we
will conjecture its complete form. The meromorphic functions which are
well  defined on the surface specified by equation (7.1) are rational
polynomials in $Q$ and
$\varsigma$. We are interested in functions that have poles at one of the
Koba-Nielsen points. Clearly, one set of such functions is
$$
{1\over (\varsigma-\varsigma_i)^p}= {1\over (\varrho_i)^{Np}}
\eqno(7.2)$$
where $\varrho_i= (\varsigma-\varsigma_i)^{{1\over N}}$.
However, one can also consider
$$
{Q^r(\varsigma-\varsigma_4)\over (\varsigma-\varsigma_3)^p}=
{a^{(3)}_0\over (\varrho_3)^{Np+r}}+{a^{(3)}_1\over
(\varrho_3)^{N(p-1)+r}}+\ldots \quad , \quad 1 \leq r<N, \ p \geq 1
\eqno(7.3)$$
where $+\ldots$ indicates less singular terms, and a similar expression with
$\varsigma_3$ and $\varsigma_4$ exchanged for the pole at $\varsigma_4$.
For poles at $\varsigma_1$ we take the functions
$$
{(\varsigma-\varsigma_2)\over Q^r(\varsigma-\varsigma_1)^p}=
{a^{(1)}_0\over (\varrho_1)^{Np+r}}+{a^{(1)}_1\over
(\varrho_1)^{N(p-1)+r}}+\ldots \quad , \quad 1 \leq r<N, \ p \geq 1
\eqno(7.4)$$
with an analogous expression for the poles at $\varsigma_2$. We note
that the functions in equations (7.3) and (7.4) are of the form
$\varrho_i^{-(Np+r)}$ times a polynomial in $\varrho_i^{N}$  and, in
particular, they contain no constant term. Furthermore these functions
vanish at the  Koba-Nielsen points where they are not singular.
We will need a good set of local
coordinates in the neighbourhood of each of the Koba-Nielsen points, and
one such choice is
$\varrho_i$. We note that
$\varsigma-\varsigma_i$ is not a good choice for reasons spelt out in the
genus one case. Equations (7.2) and (7.3) have all possible poles
at a single Koba-Nielsen point except for ${1\over (\varrho_i)^{q}},\
q=1,\ldots, N-1$. These are the expected missing poles for which no
such function exists.
\par
Using the functions of equation (7.2) in the integrated overlap equation
(2.15) we conclude that, in the high energy limit,
$$
V_\varrho\left(\alpha_{-Np}^{\mu (i)}+\sum_{j,j \not= i} {1\over
(\varsigma_j-\varsigma_i)^p}\alpha_0^{\mu (j)}\right)=0
\eqno(7.5)$$
Using superpositions of the functions in equations (7.3) and (7.4) we
also find, in the high energy limit, integrated overlap equations
of the form
$$
V_\varrho\left(\alpha_{-(Np+r)}^{\mu (i)}+c_i \ \alpha_{-r}^{\mu (i)}\right)=0
\quad , \quad r=1,\ldots, N-1,\ p\ge 1
\eqno(7.6)$$
where $c_i$ is a constant.
However,
to find the dependence of the vertex on
$\alpha_q^\mu,\ 1\le q < N$ we must use integrated overlap equations
derived from functions $\varphi$ that shift by a constant under the
periods of the surface.
\par
The dependence of the vertex on the oscillators $\alpha_{Np}^\mu,
\ \bar \alpha_{Np}^\mu, \ p\ge 1$ and $\alpha_0^\mu$ is given by

$$
\prod_{k=1}^N {}_{(k)}\langle
0|\exp\left(-\sum_{i,j}\sum_{p=1}^{\infty}\left( \alpha_{pN}^{ \mu
(i)} {1\over pN(\varsigma_j-\varsigma_i)^p}\alpha_{0 \mu}^{(j)} +
\bar \alpha_{pN}^{ \mu (i)}{1\over
pN(\bar{\varsigma_j}-\bar{\varsigma_i})^p}\alpha_{0
\mu}^{(j)}\right) \right.
$$
$$
\left. +{1\over  N}\sum_{i,j} \alpha_0^{ \mu (i)}\ln {|(
\varsigma_j- \varsigma_i)|}\alpha_{0 \mu}^{(j)}\right). \eqno(7.7)
$$
The purely momentum piece is that derived in references [2,3,4].
\par
We recall from section four that the genus one  vertex does
not depend on
$\alpha_{1}^\mu$ as a consequence of the overlap relations derived from
a non-periodic function $\varphi$. The
integrated overlap equations which are constructed from periodic
functions whose first terms have  odd  negatively moded oscillators
then imply that the vertex then does not
depend on
$\alpha_{2n+1}^\mu, n=1,2,\ldots$.  The generalisation of the first of
these  statement  to  higher genus is that
the vertex does not depend on the oscillators whose dependence
 is determined by the overlap identities which require non-periodic
functions, that is those constructed from functions $\varphi$ associated
with the missing poles on the Riemann surface  and so it does not
depend on
$\alpha_{q}^\mu\ q=1,\ldots ,N-1$. If this is true then the overlap
equations (7.6) then implies that the vertex
 does not depend on the oscillators
$\alpha_{pN+r}^\mu,\, r=1,\ldots ,N-1,\, p\ge 1$. In other words it
depends on only
$\alpha_{pN}^{ \mu},\ p\ge 1$ and as a result the vertex in equation (7.7)
is the complete  vertex for the scattering of four high energy, fixed angle,
arbitrary strings.
\par
To verify this assumption one has to find the contour integrals of
$\int V\partial x^\mu dz$ around the cuts in the surface. We will
leave this to a later paper.   However, the vertex must satisfy
the overlap equation $Vx^{\mu (i)}(\varrho_i)=Vx^{\mu
(j)}(\varrho_j)$. It is easy to verify that the vertex of equation
(7.7) does satisfy this equation. Thus if there was some
additional oscillator dependence in the vertex this factor would
have to satisfy this equation by itself; perhaps this is unlikely
as all other vertices encountered so far in string theory do not
work in this way. In the next section we will give a path integral 
derivation of this vertex.
\par

Although the vertex does not satisfy any genus
independent relations, it does have  a very simple form. As
one goes to higher and higher genus, the scattering becomes free-like as
more and more of the coefficients in the vertex become zero. Indeed, for
the contributions calculated in this paper, a  process involving a
graviton as one of the scattered particles receives only  a tree
contribution as all higher genus vertices do not contain the required
$\alpha_1$ oscillator.  Indeed, if a state
contains an oscillator $\alpha_n$ this can only get contributions from
genus g where $n=p(g+1)$ for positive integer $p$. As a result, the
contribution to the  scattering of states involving oscillators
$\alpha_{n_i}^{\mu (i)}$  occurs at most at the greatest common divisor of
the $n_i$ minus one. What is also striking is that the
coefficients that are non-zero are essentially tree-like. In this paper
we have calculated the leading term in the high energy limit at each
genus. This is a term of the form of the tachyon contribution
multiplied by $p^m$ where $m$ is the number of oscillators in all the
four particle physical states which results from the $\alpha_n\alpha_0$
term in the exponential of the vertex. However, this is not the same as
the dominant contribution taking into account the different orders of
perturbation theory. Indeed, at one loop  there is a contribution from
the full vertex which arises  from the $\alpha_n\alpha_m$
term in the exponential of the vertex. Although this involves no momenta,
it is larger than the tree level contribution referred to above because the
exponential factor at one loop is much larger than that found at tree
level.
\par
It will be interesting to see if the results found in this paper
can be used to find any symmetries hidden in the high energy scattering
amplitudes. We have in mind the symmetries proposed in reference [7] and
also those associated with higher spin theories. It will also be of
interest to generalise these results to the superstring. One expects that
 at least the Neveu-Schwarz sector of the calculation would be rather
similar. However, in this theory the tachyon is projected out in the
superstring and so   for the particles that
are actually in the theory it will be interesting to find  the analogous
high energy scattering.

\bigskip
{\bf {8. Derivation of Four Arbitrary String Scattering from the Sum over
World Sheets Approach. }}
\medskip
In this section we recover the results for four arbitrary closed string
scattering at high energy from the sum over world sheets approach to
string theory used in references [2-5]. In this approach the amplitude
is given by a Euclidean path integral of the form
$$
A=\int {\cal D} g_{\alpha\beta} {\cal D} x^\mu \exp (-S[x^\mu])  \prod _i
\int d^2\xi_i V^{(i)}(x^\mu(\xi_i),p_i).
\eqno(8.1)
$$
In this equation
$$
S[x^\mu]={1\over 4\pi \alpha^\prime}\int d^2 \xi\sqrt g g^{\alpha\beta}
\partial_\alpha x^\mu\partial_\beta x_\mu=
\int d^2 \xi {1\over 2} x^\mu (\xi) Px_\mu  (\xi),
\eqno(8.2)$$
where $P$ is the operator $P=-{1\over 2\pi \alpha^\prime} \partial_\alpha
(\sqrt g g^{\alpha\beta}\partial_\beta ) $, and
$V^{(i)}(x^\mu(\xi_i),\partial x^\mu (\xi_i),\ldots ,p_i)$ is the vertex
operator of conformal weight (1,1) corresponding to the emitted
string at the Koba-Nielsen point
$z_i$ with momentum $p_i=\sqrt {2\over \alpha^\prime}\alpha_0=\sqrt
{2\over \alpha^\prime}\bar \alpha_0$. The traditional choice of
$\alpha^\prime$ is
$\alpha^\prime=2$ for the closed string, although we will not make this
choice in this paper.  The vertex operator is  of the form
$$
V^{(i)}(x^\mu(\xi_i),\partial x^\mu (\xi_i),\ldots ,p_i)=
\sqrt g (\epsilon_{\mu\nu}
\partial^p x^\mu(\xi_i)\bar \partial^p x^\nu (\xi_i) +\ldots )\exp
(i p_i\cdot x(\xi_i))
$$
$$
\equiv U(x^\mu (\xi_i),\partial x^\mu (\xi_i),\ldots )
\exp (i p_i\cdot x(\xi_i))
\eqno(8.3)$$
where $+\ldots$ in the first line stands for the addition of terms which
contain more than two factors of $x^\mu$ which are also compatible with
the level matching condition of the closed string. In this expression
$\epsilon_{\mu\nu}$ is one of the polarisation tensors needed to specify
the physical state. Demanding that the vertex operator be of
conformal weight (1,1) implies that the
polarisation tensors are subject to certain
conditions.
\par
The path integral is over all surfaces of the genus being
considered. As we are working in the critical dimension, namely twenty-six,
the integral is invariant under the Weyl rescalings of the metric
$g_{\alpha\beta}$ and diffeomorphisms and so these must be factored out
of the path integral by incorporating ghosts. The integral is then over
the remaining variables; the moduli of the Riemann surfaces [31,34] of genus
$g$, which is a  finite dimensional space, and the Koba-Nielsen
coordinates $z_i$.
\par
We may write the path integral of equation (8.1) as
$$
A=\int {\cal D} g_{\alpha\beta}\Pi_l d^2\xi_l {\cal D} x^\mu \exp
\left(-S[x^\mu]+
\int d^2
\xi \sum_j ip_j\cdot x(\xi) \delta(\xi-\xi_j) \right)
$$
$$
\prod_k  U^{(k)}(x^\mu (\xi_k),\partial x^\mu (\xi_k),\ldots )
\eqno(8.4)$$
Completing the square in $x^\mu$ in the first exponential this expression
becomes
$$
A=\int {\cal D} g_{\alpha\beta}\Pi_l d^2\xi_l {\cal D} \tilde x^\mu \exp
\left(-S[\tilde x^\mu]\right) \exp \left(-{1\over 2}\sum_{j,k} p_j \cdot
G(\xi_j,\xi_k) p_k \right)
$$
$$\prod_k  U^{(k)}(x^\mu (\xi_k),\partial x^\mu (\xi_k),\ldots )
\eqno(8.5)$$
where  $G(\xi_j,\xi_k)$ is the Green's function on the surface which is
defined to satisfy the equation
$$
P G(\xi,\xi_k)= -{1\over 2\pi \alpha^\prime} \partial_\alpha
(\sqrt g g^{\alpha\beta}\partial_\beta ) G(\xi,\xi_k)= \delta (\xi,\xi_k)
\eqno(8.6)$$
and
$$\tilde x^\mu(\xi)=x^\mu(\xi) - i\sum_j p_j^\mu P^{-1}\delta (\xi,\xi_j)
=x^\mu(\xi)-i\sum_j p_j^\mu G(\xi,\xi_j).
\eqno(8.7)$$
The Green's function $G$ depends on the moduli of the Riemann surface.
\par
In
the high energy limit we may replace the dependence of
$U^{(i)}$, in equation (8.5), on $x^\mu(\xi_i)$ by $\hat x^\mu(\xi_i)$
where
$$
\hat x^\mu(\xi)= i \sum_j p_j^\mu G(\xi,\xi_j).
\eqno(8.8)
$$
We may now carry out the integration over $\tilde x^\mu$ to produce the
usual determinants and, following references [2-5], the high energy limit
can now be calculated by the method of steepest descent by extremising in
the moduli of the Riemann surface and the positions of the Koba-Nielsen points
$\xi_i$. The values of these  parameters are  determined by the second
exponential factor in equation (8.5) involving the momenta and the Green's
function. As such, we find that  the amplitude for string scattering is
given by that for the scattering of four tachyons, $A_{\rm tachyon}$
multiplied by the $U$ factors associated with the scattering of the
arbitrary string states, that is
$$
A=A_{\rm tachyon}\prod_k U^{(k)}(\hat x^\mu (\hat  \xi_k),\partial \hat x^\mu
(\hat \xi_k),\ldots )
\eqno(8.9)$$
where $\hat  \xi_k$ are the values of the Koba-Nielsen points at the
extremum.
The values of the moduli and Koba-Nielsen points at the extremum
fix the world-sheet to be the $\hat x^\mu$  given in  equation (8.8)
\par
It was argued in references [2-5] that in the high energy limit the
dominant Riemann surface at genus $g$ for four particle scattering could
be represented by the Riemann sphere with two  branch cuts of degree
$N=g+1$ and the four Koba-Nielsen points were at the positions $z_i$
of the branch cuts. As such,  the surface consisted of a $N$-sheeted
cover of the Riemann sphere given by
$$y^N= {(\varsigma-\varsigma_1)(\varsigma-\varsigma_2)\over
(\varsigma-\varsigma_3)(\varsigma-\varsigma_4)}.
\eqno(8.10)$$
Since the sheets are all on an equal footing the Green's function on the
surface is given by [2-5]
$$
G(\varsigma,\varsigma^\prime)= -{\alpha^\prime\over 2N}
\ln(\varsigma-\varsigma^\prime)+c.c .
\eqno(8.11)$$
\par
Using this
result  we can find  an explicit expression for  the extremal world
surface given in equation (8.8). However, it is important to remember that
$\varsigma-\varsigma_i$ is not a good coordinate at  the position of the
Koba-Nielsen
point $\varsigma_i$ where the string is emitted. A good coordinate is given by
$\varrho_i=(\varsigma-\varsigma_i)^{1\over N}$. The string surface evaluated
 in terms of the good coordinate $\varrho_i$ is given by
$$
\hat x^\mu(\varrho_i)=-{i\alpha^\prime\over 2} p^\mu_i\ln \varrho_i-
{i\alpha^\prime\over 2N}\sum_{j,j\not= i} p^\mu_j\ln
(\varrho_i^N+\varsigma_i-\varsigma_j)+c.c
\eqno(8.12)$$
Taking the derivative and expanding about $\varrho_i=0$ we find that
$$
\partial\hat x^\mu(\varrho_i)=-{i\alpha^\prime\over 2 \varrho_i} p^\mu_i-
{i\alpha^\prime\over 2}\sum_{j,j\not= i} p^\mu_j{\varrho_i^{N-1}\over
(\varrho_i^N+\varsigma_i-\varsigma_j)}+c.c
$$
$$=-{i\alpha^\prime\over 2 \varrho_i} p^\mu_i+
{i\alpha^\prime\over 2}\sum_{j,j\not= i} \sum_{p=0}^\infty
p^\mu_j{\varrho_i^{N(p+1)-1}\over  (\varsigma_j-\varsigma_i)^{p+1}}+c.c
\eqno(8.13)$$
Recalling the standard expression for the embedding field in terms of
oscillators, $\ $ $\partial
x^\mu(\varrho)=-i\sqrt{\alpha^\prime\over
2}\sum_{-\infty}^\infty\alpha_{-n}^\mu
\varrho^{n-1} $,
we conclude that in the high energy limit the oscillators
 are given by
$$
\hat \alpha_{-Np}^{\mu (i)}=-\sum_{j,j\not= i} {\hat \alpha_0^{\mu
(j)}\over  (\varsigma_j-\varsigma_i)^{p}},\ p=0,1,\ldots\ \ {\rm and }\ \
\hat \alpha_0^{\mu (i)}=\sqrt{\alpha^\prime\over 2} p_i^\mu
\eqno(8.14)$$
with all other oscillators vanishing. We note that had we not  used  the
good coordinate $\varrho_i$ we would not have recovered the
correct dependence of  $ \hat x^\mu(\varrho_i)$ on  the momentum.
\par
Given an operator of conformal weight (1,1) of the form of that in
equation (8.3) we can construct the associated physical state in the usual
way
$$\lim_{\varrho_i\to 0} V^{(i)}(x^\mu(\varrho_i) ,p_i)|0\rangle =
|\psi_i \rangle.
\eqno(8.15)$$
The positively  moded oscillators,  $\alpha_{m},\ m>0$, in
$V^{(i)}(x^\mu(\varrho_i) ,p_i)$ annihilate on the vacuum, while those that
have $\varrho_i^n, n>0$ also vanish in the limit and one is left with
a finite number of negatively moded oscilators acting on the vacuum which
now carries the momentum of the string state. We note that some
potentially divergent terms as $\varrho_i\to 0$ are absent due to the
physical state conditions, or equivalently, the demand that the operator
is of conformal weight (1,1) satisfied by  the polarisation tensors.
\par
The arbitrary four string amplitude in equation (8.9) requires us to
evaluate  the conformal operator at the  Koba-Nielsen point
$\varsigma_i$, or
$\varrho_i=0$.  This means evaluating the expression of equation (8.3) for
$\partial^p\hat x^\mu(\varrho_i)$ at $\varrho_i=0$.
However, we see from equation (8.13) that  this field contains no positively
moded oscillators and is
analytic in
$\varrho_i$, apart from some terms containing momentum that vanish in the
vertex operator due to the physical state conditions among the
polarisation tensors. Hence, it has  a smooth limit when
$\varrho_i\to 0$ which is  precisely the same as the expression of
equation (8.15)  for the physical state, except that  the oscillators
acting on the vacuum in that equation are replaced by  the hatted
oscillators which are given in terms of the momenta by  equation (8.14).
On the other hand, when evaluating the scattering amplitude using the
group theoretic approach we apply the desired physical state onto
the vertex and then evaluate it by acting with the negatively moded
oscillators to the left, that is  onto the vertex. In doing so we bring down
certain factors from the vertex which in the high energy limit contain
just the momenta. We observe that for the vertex of equation (7.7) these
are just those given in equation (8.14), see equation (7.5). As a result,
we find that the sum over world surfaces approach recovers the four
arbitrary string scattering of the group theoretic approach embodied in
the vertex of equation (7.7).
\par
The ease with which one can recover the result for four arbitrary
string scattering from the sum over world surfaces approach makes
up for some of the heuristic steps that are required, such as
the precise definition of the vertex operators used. Nonetheless
it does provide strong confirmation for the vertex of equation (7.7).

\bigskip
{\bf Acknowledgments} \par This research was supported by a PPARC
senior fellowship PPA/Y/S/2002/001/44 and by the PPARC grants
PPA/G/O/2002/00475. N.~M. would like to thank the Theory Group at CERN
for their kind hospitality during the writing up of this project.


\bigskip
{\bf {Appendix A: Representations of genus one Riemann surfaces}}
\medskip

In this Appendix, we carefully review different representations of the
torus, namely the parallelogram and the cut sphere representations,
and how they are related. We give a short review of the theory of
elliptic functions, the Weierstrass function ${\cal P}(z)$, the zeta
function $\zeta(z)$ and the sigma function $\sigma(z)$, and their
relations to the Jacobi theta functions. For more details, we refer
the reader to [24] for example.

\par
A torus can be mapped conformally to a parallelogram in the complex
plane, whose opposite sides are identified. Such a parallelogram can
be obtained by modding out the complex plane ${\bf C}$ by the
equivalence relations.
$$
z \sim z + 2 m w + 2 n w' \qquad m,n \in {\bf Z}.
\eqno(A.1)
$$
The parallelogram in
${\bf C}$ with corners at $z=0$, $2w$, $2w+2w'$ and $2w'$ is mapped
in a  one to one manner onto   the torus. We call the region inside the
parallelogram the fundamental domain. We will assume that
$w$ and
$w'$ are labelled in such a way that
${\rm Im}
\tau > 0$, where
$$
\tau = {w' \over w}.
\eqno(A.2)
$$
Of fundamental importance are the holomorphic maps from the torus onto
the Riemann sphere. These can be seen as maps $f$ from the complex
numbers to the Riemann sphere, which have to obey the periodicity
conditions:
$$
f(z+2w) = f(z) \qquad , \qquad f(z+2w') =
f(z) \qquad \forall z \in {\bf C}.
\eqno(A.3)
$$
Since we want $f$ to be onto, it must have at least one pole (because
the Riemann sphere is ${\bf C} \cup \{\infty\}$). However one can see
that $f$ cannot have only one pole; indeed if we integrate $f$ along
the circumference of the fundamental parallelogram, we will get zero
because the contributions from opposite sides will cancel. Therefore
the sum of the residues of $f$ inside the fundamental region is
zero. But $f$ can have two or more poles, or multiple poles. In fact
all such functions can be constructed from two functions: the
Weierstrass ${\cal P}$ function and its derivative ${\cal P}'(z)$. The
Weierstrass function has one double pole at $z=0$ and is holomorphic
everywhere else. It is given by
$$
{\cal P}(z) = {\cal P}(z|w,w') = {1 \over z^2} + \sum{}^\prime\left(
{1 \over (z-2mw - 2nw')^2} - {1 \over (2mw + 2nw')^2} \right),
\eqno(A.4)
$$
where the prime on the sum indicates that we exclude the point $(m,n)
= (0,0)$ from the summation. This function is even and it is
normalized so that the coefficient of the leading term in its power
expansion is one, and the constant term is zero. Its series expansion
thus starts with
$$
{\cal P}(z) = {1\over z^2} + {g_2 \over 20} z^2 +
{g_3 \over 28} z^4 + \ldots
\eqno(A.5)
$$
where $g_2$ and $g_3$ are given by
$$
g_2 = 60 \sum{}'(2mw+2nw')^{-4} \quad , \quad
g_3 = 140 \sum{}'(2mw+2nw')^{-6}.
\eqno(A.6)
$$
The derivative of the ${\cal P}$ function, ${\cal P}'(z)$, satisfies the
elliptic condition (A.3) and is odd. It is then easy to see that
$$
{\cal P}'(w) = {\cal P}'(w') = {\cal P}'(w+w') = 0.
\eqno(A.7)
$$
At these special points, it is useful to define
$$
\varsigma_1 = {\cal P}(w) \quad , \quad \varsigma_2 = {\cal P}(w+w')
\quad , \quad \varsigma_3 = {\cal P}(w').
\eqno(A.8)
$$
Using this, (A.7) can be written
$$
{\cal P}'(\varsigma_i) = 0 \qquad i=1,2,3.
\eqno(A.9)
$$
One of the fundamental properties of ${\cal P}(z)$ is that its
derivative satisfies the algebraic equation
$$
{\cal P}'^2(z) = 4 {\cal P}^3(z) - g_2 {\cal P}(z) - g_3 = 4 ({\cal
P}(z)-\varsigma_1) ({\cal P}(z)-\varsigma_2) ({\cal P}(z)-\varsigma_3)
\eqno(A.10)
$$
where, in the last equality, we have used (A.9). A way to see why we
  have such a relation is to note that both sides of this equation
  have the same principal part (the part of the series expansion with
  negative powers), their difference is therefore a holomorphic
  function from the torus to the sphere. It maps the torus to a set
  that is compact and doesn't contain infinity, hence bounded. From
  Liouville's theorem, we deduce that this function is a constant. We
  may then use (A.4) to show that this constant is zero. It follows
  from (A.10) that
$$
{\cal P}''(z) = 6 {\cal P}^2(z) - {1 \over 2} g_2 \quad , \quad
{\cal P}'''(z) = 12 {\cal P}(z) {\cal P}'(z).
\eqno(A.11)
$$
And by induction we conclude that ${\cal P}^{(2n-2)}(z)$ and ${\cal
  P}^{(2n+1)}(z)/{\cal P}'(z)$ are polynomials of degree $n$ in ${\cal
  P}(z)$.

\bigskip
\par
We can now describe the cut sphere representation of the torus. We
consider the mapping
$$
\varsigma = {\cal P}(z)
\eqno(A.12)
$$
from the torus to the Riemann sphere (parameterized by
$\varsigma$). Under this holomorphic mapping, every point of the
sphere is reached exactly twice (counting multiplicity). The
points which are reached once but have multiplicity two are those
at which ${\cal P}'$ is zero, these are $\varsigma_1$,
$\varsigma_2$ and $\varsigma_3$. And $\varsigma_4 \equiv \infty$
also has multiplicity two, as can be seen from (A.5). We can
therefore make the mapping (A.12) one-to-one by considering,
instead of the sphere, the Riemann surface obtained by taking two
copies of the Riemann sphere, making cuts on each of these two
sheets between $\varsigma_1$ and $\varsigma_2$ and between
$\varsigma_3$ and $\infty$, and then gluing the two sheets
together in such a way that when we go through a cut, we actually
just go from one sheet to another (see figure 2). We will call
this surface the cut sphere. It was constructed so as to make the
holomorphic mapping (A.12) onto and one-to-one. \break

\input epsf
\medskip
\epsfbox{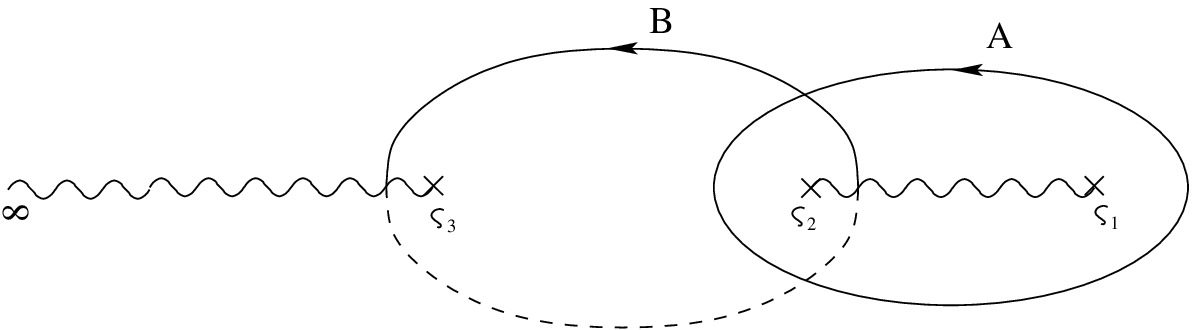}
\medskip
{\bf Figure 2}: The cut sphere. The two cuts, drawn as wavy lines,
are between $\varsigma_1$ and $\varsigma_2$ and between $\varsigma_3$
and $\infty$. We have drawn two independent nontrivial closed
curves. The curve $A$ doesn't cross any cut, so it remains on the top
sheet. The curve $B$ however, crosses the cut between $\varsigma_3$
and $\infty$, where it goes to the lower sheet (it is then drawn with
a dashed line). It then crosses the other cut and comes back to the
top sheet.
\medskip

\par
We now want to find good local coordinates on the cut sphere. First,
at a regular point $\varsigma_0$ where ${\cal P}'(z) \neq 0$, we have
$$
\varsigma = \varsigma_0 + a_1 z + a_2 z^2 + \ldots \qquad a_1 \neq 0
\eqno(A.13)
$$
and therefore $\varsigma-\varsigma_0$ is a good local coordinate.
Now, at the points $\varsigma_i$, where ${\cal P}^\prime(z) = 0$,
we have
$$
\varsigma = \varsigma_i + 0 \cdot z + a_2 z^2 + \ldots \qquad a_2
\neq 0
\eqno(A.14)
$$
therefore
$$
z \sim {1 \over \sqrt{a_2}} \sqrt{\varsigma-\varsigma_i},
\eqno(A.15)
$$
hence a good local coordinate is
$$
\varrho_i = \sqrt{\varsigma-\varsigma_i}.
\eqno(A.16)
$$

\par
Now we would like to find the reciprocal of (A.12). For this we start
by defining
$$
Q(\varsigma) = -2 \sqrt{(\varsigma-\varsigma_1) (\varsigma-\varsigma_2)
(\varsigma-\varsigma_3)}.
\eqno(A.17)
$$
Let us consider the integral
$$
I = \int_{\infty}^{{\cal P}(z)}{\varsigma' \over Q(\varsigma')},
$$
and let us make the change of variable $\varsigma' = {\cal P}(z')$,
then we see from (A.10) that $Q(\varsigma') = {\cal
P}'(z')$. Therefore
$$
I = \int_0^zdz' = z,
$$
and we thus find the inversion formula
$$
z = \int_{\infty}^\varsigma{d\varsigma' \over Q(\varsigma')},
\eqno(A.18)
$$
where $\varsigma = {\cal P}(z)$. Note that we haven't defined the path
of integration. And since there are nontrivial closed paths going
around the cuts (like the path $A$ of Figure 2) or going through
both cuts (like the path $B$), it seems that which path we choose will
matter. What happens actually is that by going around a closed path,
the integral picks up an integer combination of the periods
$$
\oint{d\varsigma' \over Q(\varsigma')} = 2mw + 2nw',
\eqno(A.19)
$$
and therefore $z$ is unaffected due to the identifications (A.1).  We
can actually be more precise, looking back at Figure 2 and the
contours $A$ and $B$, we can see that
$$
\oint_A{d\varsigma' \over Q(\varsigma')} = 2w' \quad , \quad
\oint_B{d\varsigma' \over Q(\varsigma')} = -2w.
\eqno(A.20)
$$
Indeed, in the $z$ plane, the contours $A$ and $B$ go respectively
from $w$ to $w+2w'$ and from $2w+w'$ to $w'$, winding only once on the
torus.  The result then follows from (A.18).

\bigskip
\par
We now consider other useful functions, which we are using in this
paper. First, the zeta function is minus the integral of the ${\cal
P}$ function
$$
{\cal P}(z) = -\zeta'(z).
\eqno(A.21)
$$
This defines $\zeta(z)$ up to a constant, which we choose so that the
series expansion starts like
$$
\zeta(z) = {1 \over z} - {g_2 \over 60} z^3 -
{g_3 \over 140} z^5 - \ldots
\eqno(A.22)
$$
The zeta function is not doubly periodic, but as can be seen from
(A.21), under translation by a period it changes only by a constant
(independent of $z$)
$$
\zeta(z+2w) = \zeta(z) + 2 \eta \quad , \quad \zeta(z+2w')
= \zeta(z) + 2 \eta'.
\eqno(A.23)
$$
The two constants $\eta$ and $\eta'$ are the values of the zeta
functions at the half periods:
$$
\eta = \zeta(w) \quad , \quad \eta' = \zeta(w').
\eqno(A.24)
$$
By integrating $\zeta(z)$ around a fundamental parallelogram, one
finds the Legendre's relation
$$
\eta w' - \eta' w = {i \pi \over 2}.
\eqno(A.25)
$$
Another useful function is the $\sigma$ function which is defined by
$$
\zeta(z) = {d \over dz} \ln \sigma(z),
\eqno(A.26)
$$
and normalized such that
$$
\sigma(z) = z - {g_2 \over 240} z^5 - {g_3 \over 840} z^7 - \ldots
\eqno(A.27)
$$
Again, this is not a doubly periodic function; under a shift by a
period, it transforms as
$$
\sigma(z+2w) = -\sigma(z) e^{2 \eta(z+w)} \quad , \quad
\sigma(z+2w') = -\sigma(z) e^{2 \eta'(z+w')}. \eqno(A.28)
$$

\bigskip
\par
It is useful to relate the previous functions to the four Jacobi theta
functions. We recall here their definitions as infinite sums
$$
\theta_1(v) = \theta_1(v|\tau) = i \sum_{n=-\infty}^\infty(-1)^n
q^{(n-{1\over2})^2} e^{i \pi (2n-1) v}
$$
$$
\theta_2(v) = \theta_2(v|\tau) = \sum_{n=-\infty}^\infty
q^{(n-{1\over2})^2} e^{i \pi (2n-1) v}
$$
$$
\theta_3(v) = \theta_3(v|\tau) = \sum_{n=-\infty}^\infty
q^{n^2} e^{i \pi 2n v}
$$
$$
\theta_4(v) = \theta_4(v|\tau) = \sum_{n=-\infty}^\infty
(-1)^n q^{n^2} e^{i \pi 2n v},
\eqno(A.29)
$$
where $q = e^{i \pi \tau} = e^{i \pi {w' \over w}}$. They can also be
expressed as infinite products
$$
\theta_1(v) = 2 f_0(q^2) q^{1 \over 4} \sin \pi v \prod_{n=1}^\infty
(1-2q^{2n} \cos 2\pi v + q^{4n})
$$
$$
\theta_2(v) = 2 f_0(q^2) q^{1 \over 4} \cos \pi v \prod_{n=1}^\infty
(1+2q^{2n} \cos 2\pi v + q^{4n})
$$
$$
\theta_3(v) = f_0(q^2) \prod_{n=1}^\infty
(1+2q^{2n-1} \cos 2\pi v + q^{4n-2})
$$
$$
\theta_4(v) = f_0(q^2) \prod_{n=1}^\infty
(1-2q^{2n-1} \cos 2\pi v + q^{4n-2}),
\eqno(A.30)
$$
where we have defined
$$
f_0(q^2) = \prod_{n=1}^\infty(1-q^{2n}).
\eqno(A.31)
$$
In all these and following expressions, $v$ is related to $z$ by
$$
v = {z \over 2w}.
\eqno(A.32)
$$
In other words, the periods in the $v$ plane are $1$ and $\tau$. The
theta functions obey the following periodicity equations.
$$
\theta_1(v+1) = - \theta_1(v) \quad , \quad \theta_1(v+\tau) =
- e^{-i \pi (2v+\tau)} \theta_1(v)
$$
$$
\theta_2(v+1) = - \theta_2(v) \quad , \quad \theta_2(v+\tau) =
e^{-i \pi (2v+\tau)} \theta_2(v)
$$
$$
\theta_3(v+1) = \theta_3(v) \quad , \quad \theta_3(v+\tau) =
e^{-i \pi (2v+\tau)} \theta_3(v)
$$
$$
\theta_4(v+1) = \theta_4(v) \quad , \quad \theta_4(v+\tau) =
-e^{-i \pi (2v+\tau)} \theta_4(v)
\eqno(A.33)
$$
The theta functions are related to each other by translations by half
periods. For example for $\theta_1$ we have
$$
\theta_1(v+{1\over2}) = \theta_2(v) \quad , \quad
\theta_1(v+{\tau \over 2}) = i e^{-i \pi(v+{\tau\over4})} \theta_4(v),
\eqno(A.34)
$$
and we have similar relations for the other theta functions. We can
express all the functions defined above from the Weierstrass
functions, in terms of theta functions, For example we have
$$
\zeta(z) = {\eta \over w}z + {1 \over 2w}
{\theta_1'(v) \over \theta_1(v)}
\eqno(A.35)
$$
$$
\sigma(z) = 2w \exp\left({\eta z^2 \over 2w}\right)
{\theta_1(v) \over \theta_1'(0)}
\eqno(A.36)
$$
and
$$
\eta = -{1 \over 12 w} {\theta_1'''(0) \over
\theta_1'(0)} \quad , \quad
\eta' = -{i \pi \over 2 w} -{\tau \over 12 w} {\theta_1'''(0)
\over \theta_1'(0)}.
\eqno(A.37)
$$
The theta functions obey many more relations; here we have only
written the ones that we are using in this paper. For a more detailed
introduction, see for example [24].

\bigskip
{\bf References}
\medskip
\item{[1]} V.~Alessandrini, D.~Amati and B.~Morel, {\it The Asymptotic
Behaviour of the Dual Pomeron Amplitude}, Il Nuovo Cimento, vol7 (1972)
797.

\item{[2]} D.~J.~Gross and P.~F.~Mende,
  {\it The High-Energy Behavior Of String Scattering Amplitudes},
  Phys.\ Lett.\ B {\bf 197} (1987) 129.

\item{[3]}  D.~J.~Gross and P.~F.~Mende,
  {\it String Theory Beyond The Planck Scale},
  Nucl.\ Phys.\ B {\bf 303} (1988) 407.

\item{[4]} D.~J.~Gross and J.~L.~Manes,
  {\it The High-Energy Behavior Of Open String Scattering},
  Nucl.\ Phys.\ B {\bf 326} (1989) 73.

\item{[5]} D.~J.~Gross,
  {\it High-Energy Symmetries Of String Theory},
  Phys.\ Rev.\ Lett.\  {\bf 60} (1988) 1229.

\item{[6]} C.~T.~Chan and J.~C.~Lee,
  {\it Stringy symmetries and their high-energy limits},
  Phys.\ Lett.\ B {\bf 611} (2005) 193
  [arXiv:hep-th/0312226];
  C.~T.~Chan and J.~C.~Lee,
  {\it Zero-norm states and high-energy symmetries of string theory},
  Nucl.\ Phys.\ B {\bf 690} (2004) 3
  [arXiv:hep-th/0401133];
  C.~T.~Chan, P.~M.~Ho and J.~C.~Lee,
  {\it Ward identities and high-energy scattering amplitudes in string theory},
  Nucl.\ Phys.\ B {\bf 708} (2005) 99
  [arXiv:hep-th/0410194];
  C.~T.~Chan, P.~M.~Ho, J.~C.~Lee, S.~Teraguchi and Y.~Yang,
  {\it Solving all 4-point correlation functions for bosonic open string theory
in
  the high energy limit},
  arXiv:hep-th/0504138;
  C.~T.~Chan, P.~M.~Ho, J.~C.~Lee, S.~Teraguchi and Y.~Yang,
  {\it High-energy zero-norm states and symmetries of string theory},
  arXiv:hep-th/0505035.

\item{[7]} P.~West, $E_{11}$ and M Theory, Class.Quant.Grav.
18 (2001) 4443-4460, hep-th/0104081.

\item{[8]} V.~Alessandrini, D.~Amati, M.~Le Bellac and D.~Olive, {\it The
Operator Approach to Dual Multiparticle Theory}, Phys. Reports {\bf 6}
(1971) 269; D. Ebert and H. Otto, {\it A survey on Dual Tree and Loop
Amplitudes}, Zeitschrift Fortschritte der Physik, {\bf 25} (1977) 203.

\item{[9]} A.~Neveu and P.~C.~West, {\it Conformal Mappings and the Three
String Bosonic Vertex}; Phys. Lett. {\bf B179} 235 (1986).

\item{[10]} A.~Neveu and P.~C.~West, {\it The Cyclic Symmetric Vertex for
Three Arbitrary Neveu-Schwartz Strings}; Phys. Lett. {\bf B180}
34 (1986).

\item{[11]}  A.~Neveu and P.~C.~West,
  {\it Group Theoretic Approach To The Perturbative String S Matrix},
  Phys.\ Lett.\ B {\bf 193} (1987) 187.

\item{[12]} A.~Neveu and P.~C.~West,
  {\it A Group Theoretic Method For String Loop Diagram},
  Phys.\ Lett.\ B {\bf 194} (1987) 200.

\item{[13]} A.~Neveu and P.~C.~West,
  {\it Group Theoretic Approach To The Open Bosonic String Multiloop S
Matrix},
  Commun.\ Math.\ Phys.\  {\bf 114} (1988) 613.

\item{[14]} A.~Neveu and P.~C.~West,
  {\it Cycling, Twisting And Sewing In The Group Theoretic Approach To
Strings},
  Commun.\ Math.\ Phys.\  {\bf 119} (1988) 585.

\item{[15]} P.~West and M.~Freeman, {\it Ghost Vertices for the Bosonic
String using the Group Theoretical Approach to String Theory} ;
Phys. Lett. {\bf B205} 30 (1988).

\item{[16]} P.~West, {\it Multiloop Ghost Vertices and the Determination
of the Multiloop Measure}; Phys. Lett. {\bf B205} 38 (1988).

\item{[17]}P.~West, {\it String Vertices and Induced Representations};
Nucl. Phys. {\bf B320} 103 (1989).

\item{[18]} P.~C.~West,
  {\it A Brief Review Of The Group Theoretic Approach To String Theory},
A Brief Review of the Group Theoretic Approach to String Theory,
in "Conformal Field Theories and Related Topics", Proceedings of
Third Annecy Meeting on Theoretical Physics, LAPP, Annecy
le Vieux, France, Nucl. Phys. B (Proc. Suppl) {\bf 5B} (1988) 217, edited
by P. Bin\"utruy, P. Sorba and R. Stora, North Holland (1988); and
"Ninth Workshop on Grand Unification"; edited by R. Barloutaud, World
Scientific (1988); and
 "1988 Electroweak Interactions and Unified
Theories", Proceedings of the XXIIIrd Rencontre de Moriond, edited by J.
Tran Thanh Van, published by Edition Fronti\'eres (1988). CERN-TH-5059-88

\item{[19]} A.~Neveu and P.~C.~West,
  {\it Neveu-Schwarz Excited String Scattering: A Superconformal Group
Computation},
  Phys.\ Lett.\ B {\bf 200} (1988) 275;
 A.~Neveu and P.~C.~West,
  {\it Group Theoretic Approach To The Superstring And Its Supermoduli},
  Nucl.\ Phys.\ B {\bf 311} (1988) 79.

\item{[20]} M.~D.~Freeman and P.~C.~West,
  {\it Ramond String Scattering In The Group Theoretic Approach To String
Theory},
  Phys.\ Lett.\ B {\bf 217} (1989) 259.

\item {[21]} A.~Neveu and P.~C.~West, {\it Symmetries of the Interacting
Gauge Covariant  Bosonic String} ; Nucl. Phys. {\bf B278}
601 (1986).

\item{[22]} R.~W.~Gebert, H.~Nicolai and P.~C.~West,
  {\it Multistring vertices and hyperbolic Kac-Moody algebras},
  Int.\ J.\ Mod.\ Phys.\ A {\bf 11} (1996) 429
  [arXiv:hep-th/9505106].

\item{[23]} L.~Alvarez-Gaume, C.~Gomez, G.~Moore and C.~Vafa, {\it
Strings in the Operator Formalism}, Nucl. Phys. {\bf B303} (1988) 455.

\item{[24]} A.~Erdelyi, {\it Bateman Manuscript Project vol II},
McGraw-Hill, (1953).

\item{[25]} L.~Alvarez-Gaume, C.~Gomez and C.~Reina, {\it Loop Groups,
Grassmanians and String Theory}, Phys.\ Lett.\ B {\bf 190} (1987) 55;
C. Vafa, {\it Operator Formulation on Riemann Surfaces},
Phys.\ Lett.\ B {\bf 190} (1987) 47.

\item{[26]} D.~Amati, C.~Bouchiat, J.~Gervais,
{\it On the building of dual diagrams from unitarity}, Lett. Nuovo cimento
{\bf 2} (1969) 399; K. Bardacki, M. Halpern and J. Shapiro,
{\it Unitary Closed Loops in Reggeized Feynman theory}, Phys. Rev. {\bf
D185} (1969) 1910.

\item{[27]} D.~Gross, A.~Neveu, J.~Scherk and J.~Schwarz,
{\it Renormalisation and unitarity in the dual-resonance model}, Phys.
Rev. {\bf D2} (1970) 697.

\item{[28]} A.~Neveu and J.~Scherk, {\it Parameter-free regularization of
one loop unitary diagram}, Phys. Rev. {\bf D1} (1970) 311.

\item{[29]}  D.~Gross, and J.~Schwarz,
{\it Basic operators of the dual resonance model}, Nucl. Phys, {\bf B23}
(1970) 333.

\item{[30]}
  V.~Alessandrini,
  {\it A General Approach To Dual Multiloop Diagrams},
  Nuovo Cim.\ A {\bf 2} (1971) 321.

\item{[31]}
  V.~Alessandrini and D.~Amati,
  {\it Properties Of Dual Multiloop Amplitudes},
  Nuovo Cim.\ A {\bf 4} (1971) 793.

\item{[32]}  E.~Cremmer, Nucl. Phys, {\bf B31} (1971) 477.

\item{[33]} L.~Brink and D.~Olive, Nucl. Phys, {\bf B58}
(1973) 237;  L.~Brink and D.~Olive, Nucl. Phys, {\bf B56} (1973) 253.

\item{[34]} O.~Alvarez,
  {\it Theory Of Strings With Boundaries: Fluctuations, Topology, And Quantum
  Geometry},
  Nucl.\ Phys.\ B {\bf 216} (1983) 125;
E.~D'Hoker and D.~H.~Phong,
  {\it Multiloop Amplitudes For The Bosonic Polyakov String},
  Nucl.\ Phys.\ B {\bf 269} (1986) 205;
G.~W.~Moore and P.~Nelson,
  {\it Absence Of Nonlocal Anomalies In The Polyakov String},
  Nucl.\ Phys.\ B {\bf 266} (1986) 58;
J.~Polchinski,
  {\it Evaluation Of The One Loop String Path Integral},
  Commun.\ Math.\ Phys.\  {\bf 104} (1986) 37.

\end